%% file: Overview_AV2_Transform_Entropy_Coding.tex
\documentclass[conference]{IEEEtran}
\IEEEoverridecommandlockouts

\usepackage{cite}
\usepackage{amsmath,amssymb,amsfonts}
\usepackage{algorithmic}
\usepackage{graphicx}
\usepackage{multirow}
\usepackage{subfig}
\usepackage{graphicx}
\usepackage{gensymb}
\usepackage{subcaption}
\usepackage{array}
\usepackage{amsmath}
\usepackage{textcomp}
\usepackage{tabularx}
\usepackage{booktabs}   
\usepackage{xcolor}
\usepackage{makecell}
\def\BibTeX{{\rm B\kern-.05em{\sc i\kern-.025em b}\kern-.08em
    T\kern-.1667em\lower.7ex\hbox{E}\kern-.125emX}}
\begin{document}

\title{Transform and Entropy Coding in AV2}
\author{
    Alican Nalci\IEEEauthorrefmark{1},
    Hilmi E. Egilmez\IEEEauthorrefmark{2},
    Madhu P. Krishnan\IEEEauthorrefmark{3},
    Keng-Shih Lu \IEEEauthorrefmark{4},
    Joe Young \IEEEauthorrefmark{4},  \\
    Debargha Mukherjee \IEEEauthorrefmark{4},
    Lin Zheng \IEEEauthorrefmark{4},
    Jingning Han \IEEEauthorrefmark{4},
    Joel Sole  \IEEEauthorrefmark{5},
    Xiaoqing Zhu \IEEEauthorrefmark{5},
    Xin Zhao \IEEEauthorrefmark{2},\\
    Tianqi Liu \IEEEauthorrefmark{3},
    Liang Zhao \IEEEauthorrefmark{3},
    Todd Nguyen  \IEEEauthorrefmark{4},
    Urvang Joshi \IEEEauthorrefmark{4},
    Kruthika Koratti Sivakumar \IEEEauthorrefmark{4}, \\
    Luhang Xu \IEEEauthorrefmark{6},
    Zhijun Lei\IEEEauthorrefmark{1},
    Van Luong Pham \IEEEauthorrefmark{2},
    Yue Yu \IEEEauthorrefmark{6},
    Aki Kuusela \IEEEauthorrefmark{2},\\
    Minhua Zhou \IEEEauthorrefmark{7},
    Andrey Norkin \IEEEauthorrefmark{8},
    Adrian Grange \IEEEauthorrefmark{4} \\
    ~~\\
    \IEEEauthorrefmark{1}Meta,
    \IEEEauthorrefmark{2}Apple,
    \IEEEauthorrefmark{3}Tencent,
    \IEEEauthorrefmark{4}Google,
    \IEEEauthorrefmark{5}Netflix,
    \IEEEauthorrefmark{6}OPPO,
    \IEEEauthorrefmark{7}Broadcom,
    \IEEEauthorrefmark{8}Nvidia
}

\maketitle
\pagestyle{plain}
\thispagestyle{plain}

\begin{abstract}
AV2 is the successor to the AV1 video coding format developed by the Alliance for Open Media (AOMedia).
Its primary objective is to deliver substantial compression gains and subjective quality improvements while maintaining low-complexity encoder and decoder operations.
This paper describes the transform, quantization and entropy coding design in AV2, including redesigned transform kernels and data-driven transforms, expanded transform partitioning, and a mode \& coefficient dependent transform signaling.
AV2 introduces several new coding tools including Intra/Inter Secondary Transforms (IST), Trellis Coded Quantization (TCQ), Adaptive Transform Coding (ATC), Probability Adaptation Rate Adjustment (PARA), Forward Skip Coding (FSC), Cross Chroma Component Transforms (CCTX), Parity Hiding (PH) tools and improved lossless coding.
These advances enable AV2 to deliver the highest quality video experience for video applications at a significantly reduced bitrate.
\end{abstract}

\section{Introduction}
Global demand for digital video has surged over the last decade, making it the dominant driver of internet bandwidth consumption and a major source of compute and storage load in content delivery systems \cite{ericsson-2025, itu-usage, cisco-annual}.
This demand is fueled by applications such as video streaming and transcoding, video-on-demand (VOD), real-time communications (RTC), and emerging domains including immersive AR/VR, AI-generated content, and cloud gaming \cite{kumar2024cloud, bukhari2023video, meta-reels}.
The widespread adoption of 4K and 8K displays, higher frame rates, high dynamic range (HDR), and volumetric or interactive content is further expected to increase the demands on video compression technology \cite{chiariotti2021survey}.
These trends place a significant strain on delivery infrastructures, where bandwidth and storage remain costly resources.

The Alliance for Open Media (AOMedia) has developed the AV2 video codec as the successor to AV1~\cite{av2-spec}.
Released in 2018, AV1 delivered 30\% bitrate savings over VP9 at equivalent quality and has since been deployed widely across web platforms and devices \cite{han2021av1}.
AV2 extends this trajectory by providing an additional 30\% bitrate reduction over AV1.
The design of AV2 emphasizes reduced implementation and hardware cost ensuring practicality for both software and hardware deployments.

Among the key advances in AV2 are improvements to transform and entropy coding, which determines compression efficiency by advanced energy compaction and a more efficient signaling of coefficients and side information.

On the transform side, AV2 introduces redesigned kernels for primary transforms such as the discrete cosine transform (DCT), discrete sine transform (DST), and asymmetric DST (ADST).
Data-driven transforms (DDTs) are introduced to improve coding efficiency by aligning kernels with empirical residual statistics.
The transform partitioning framework is expanded with additional partitioning types for finer transform block (TB) splits.

AV2 innovations further include intra/inter secondary transforms (IST), which improve energy compaction for coefficients obtained after a primary transform with learned kernels, and cross-chroma component transforms (CCTX) that exploit coefficient correlation across chroma planes after a primary chroma transform. A coefficient and mode-dependent transform signaling scheme is also introduced which includes DC-based transform signaling (DCTX) and the mode-dependent transform derivation (MDTX) to reduce side-information signaling associated with the transform type signaling.

In terms of quantization and coefficient coding, AV2 improves the design of the default scalar quantizer from AV1 and builds on top it to enable trellis-coded quantization (TCQ) to achieve higher compression efficiency with modest additional complexity.
AV2 retains the multi-symbol arithmetic coding (MS-AC) engine of AV1 and further improves its efficiency through several new entropy coding tools.
The probability adaptation rate adjustment (PARA) tool provides an optimized probability adaptation for MS-AC engine by adjusting the adaptation rate per syntax.
Adaptive transform coding (ATC) improves the coefficient context modeling and unifies coefficient scanning rules over AV1.
Adaptive truncated Rice coding enhances the representation of large coefficient magnitudes,
and forward skip coding (FSC) enables compact representation of directly coded residuals enhancing coding efficiency for screen-content and denser residuals.
Lastly, a new parity hiding (PH) tool reduces signaling overhead associated with the DC coefficient magnitudes, and can bring coding gains in the absence of TCQ.

These coding tools provide significant efficiency gains across both natural video and screen content while preserving high throughput and low-latency operation.
This paper provides an overview of the transform and entropy coding design in AV2.
Section~\ref{sec:av1-overview} reviews the AV1 framework to establish the legacy AV1 design.
Section~\ref{sec:av2-overview} details the transform and entropy coding methods of AV2.
Section~\ref{sec:coding-gains} summarizes compression performance per coding tool.

\section{Background on AV1 Transform Coding}\label{sec:av1-overview}
\subsection{Transform Block Partitioning in AV1}\label{av1:tx_part}
Transform partitioning splits a larger coding block (CB) into smaller transform blocks (TB) for more precise handling of transform type selection and coefficient coding.
This allows the encoder to adapt the transform size to diverse residual structures and improve energy compaction through rate-distortion (RD) guided selection.

AV1 supports TB sizes ranging from $4 \times 4$ up to $64 \times 64$, including both square and rectangular configurations such as $N \times N/2$, $N \times N/4$, $N/2 \times N$, and $N/4 \times N$ \cite{han2021av1}.
Transform partitioning is applied recursively to split a CB into multiple TBs.

For intra-coded blocks, a uniform TB size is enforced such that all TBs within the same CB share the same size after a split.
This is achieved with recursive quad-tree transform partitioning for intra where the maximum TB size equals the CB size and luma TBs may again be recursively partitioned up to two recursive levels.
For inter-coded blocks, the initial TB dimensions matches the CB size and up to two recursive partition levels are allowed for the luma component.
Partition rules apply to both square and rectangular transforms permitting splits down to $N \times N/4$ or $N/4 \times N$.

\begin{table}[!t]
\centering
\caption{AV1 and AV2 Primary Transform Types}
\label{tab:tx-types-av2}
\renewcommand{\arraystretch}{1}
\begin{tabular}{|c|c|c|c|c|}
\hline
\textbf{id} & \textbf{Transform Type} & \textbf{Vertical} & \textbf{Horizontal} & \textbf{Direction} \\
\hline
0 & DCT\_DCT           & DCT        & DCT        & 2D \\
1 & ADST\_DCT          & ADST       & DCT        & 2D \\
2 & DCT\_ADST          & DCT        & ADST       & 2D \\
3 & ADST\_ADST         & ADST       & ADST       & 2D \\
4 & FLIPADST\_DCT      & FLIPADST   & DCT        & 2D \\
5 & DCT\_FLIPADST      & DCT        & FLIPADST   & 2D \\
6 & FLIPADST\_FLIPADST & FLIPADST   & FLIPADST   & 2D \\
7 & ADST\_FLIPADST     & ADST       & FLIPADST   & 2D \\
8 & FLIPADST\_ADST     & FLIPADST   & ADST       & 2D \\
9 & IDTX               & Identity   & Identity   & 2D \\
\hline
10 & V\_DCT             & DCT        & Identity   & 1D \\
11 & H\_DCT             & Identity   & DCT        & 1D \\
12 & V\_ADST            & ADST       & Identity   & 1D \\
13 & H\_ADST            & Identity   & ADST       & 1D \\
14 & V\_FLIPADST        & FLIPADST   & Identity   & 1D \\
15 & H\_FLIPADST        & Identity   & FLIPADST   & 1D \\
\hline
\end{tabular}
\end{table}

\subsection{Primary Transforms in AV1}
AV1 uses a family of separable two-dimensional (2D) trigonometric transforms constructed from combinations of one dimensional (1D) transform kernels.
The available 1D transform kernels include the Discrete Cosine Transform (DCT), the Asymmetric Discrete Sine Transform (ADST), the flipped version of the ADST kernel (FLIPADST) and the identity transform (IDTX).
By pairing horizontal and vertical 1D transforms AV1 defines up to 16 distinct 2D transforms as summarized in Table \ref{tab:tx-types-av2}.

The DCT transform is effective for compacting smoother residuals and homogeneous regions, whereas ADST and FLIPADST transforms provide better energy compaction for directional edges often encountered in inter predicted residuals.
The IDTX is useful for preserving sharp discontinuities such as those encountered in screen content \cite{chen2019av1tools,han2021av1}.

\subsection{Transform Type Signaling in AV1}
AV1 signals the transform type for each TB from a transform set $\mathrm{TXSET}$ that depends on both the TB size and the intra or inter prediction mode.
The list of transform sets, the transform types contained in each set, and the mapping of TB size and prediction type to each set are summarized in Table \ref{tab:av1-tx-sets-a} and Table \ref{tab:av1-tx-sets-b}.
This approach limits the bitstream overhead associated with transform type signaling by choosing a smaller number of symbols to code per TB.

\begin{table*}[!t]
\centering
\renewcommand{\arraystretch}{1.25}
\subfloat[AV1 transform sets and the corresponding transform types contained withing each set.\label{tab:av1-tx-sets-a}]{
\begin{tabular}{l|l|c}
\toprule
\textbf{Transform Set ($\mathrm{TXSET}$)} & \textbf{Transform Types ($tx\_type$)} & \textbf{\#} \\
\midrule
\texttt{SET\_DCTONLY} & \{\texttt{DCT\_DCT}\} & 1 \\
\texttt{SET\_DCT\_IDTX} & \{\texttt{DCT\_DCT}, \texttt{IDTX}\} & 2 \\
\texttt{SET\_DTT4\_IDTX} & \{\texttt{DCT\_DCT}, \texttt{DCT\_ADST}, \texttt{ADST\_DCT}, \texttt{ADST\_ADST}, \texttt{IDTX}\} & 5 \\
\texttt{SET\_DTT4\_IDTX\_1DDCT} & \texttt{SET\_DTT4\_IDTX} $\cup$ \{\texttt{V\_DCT}, \texttt{H\_DCT}\} & 7 \\
\texttt{SET\_DTT9\_IDTX\_1DDCT} & \texttt{SET\_ALL16} exclude \{\texttt{V\_ADST}, \texttt{H\_ADST}, \texttt{V\_FLIPADST}, \texttt{H\_FLIPADST} \} & 12 \\
\texttt{SET\_ALL16} & All candidates in Table \ref{tab:tx-types-av2} & 16 \\
\bottomrule
\end{tabular}
}
\vspace{1em}
\subfloat[Mapping of TB size and prediction type to transform sets.\label{tab:av1-tx-sets-b}]{
\begin{tabular}{c|c|c}
\toprule
\textbf{TB Size (min,max)} & \textbf{Intra-coded TB} & \textbf{Inter-coded TB} \\
\midrule
$\min(w,h)=4$, $\max(w,h)<32$  & \texttt{SET\_DTT4\_IDTX\_1DDCT} & \texttt{SET\_ALL16} \\
$\min(w,h)=8$, $\max(w,h)<32$  & \texttt{SET\_DTT4\_IDTX\_1DDCT} & \texttt{SET\_ALL16} \\
$\min(w,h)=16$, $\max(w,h)<32$ & \texttt{SET\_DTT4\_IDTX}        & \texttt{SET\_DTT9\_IDTX\_1DDCT} \\
$\max(w,h)=32$                  & \texttt{SET\_DCTONLY}          & \texttt{SET\_DCT\_IDTX} \\
$\max(w,h)=64$                  & \texttt{SET\_DCTONLY}          & \texttt{SET\_DCTONLY} \\
\bottomrule
\end{tabular}
}
\caption{AV1 transform sets and size/prediction mapping used for transform type signaling. \subref{tab:av1-tx-sets-a} lists the available transform sets and their members, and
\subref{tab:av1-tx-sets-b} shows the mapping from TB size and prediction type to these sets.}
\label{tab:av1-tx-sets}
\end{table*}

For luma TBs, the selected transform type $tx\_type$ $\in$ $\mathrm{TXSET}$ is entropy coded with syntax elements \texttt{intra\_tx\_type} or \texttt{inter\_tx\_type} for intra and inter coded blocks, respectively.

As shown in Table \ref{tab:av1-tx-sets-a}, the number of coded symbols for a TB corresponds to the number of available transforms in the respective transform set, whether the TB is intra or inter-coded and further depends on the TB size constraints such as minimum and maximum width ($w$) and height ($h$).


Smaller inter TBs allow the full set of 16 transform types, i.e., \texttt{SET\_ALL16}, while the number of transform candidates decreases with increasing TB size.
For intra TBs, the maximum set size is seven, with \texttt{SET\_DTT4\_IDTX\_1DDCT} applied to the smallest blocks.

Chroma TBs do not signal a separate transform type.
Instead, intra chroma blocks infer the transform type from the intra prediction mode, whereas inter chroma blocks inherit the transform type from the collocated luma TB.

\begin{figure}[!b]
\centering
\includegraphics[width=0.40\textwidth]{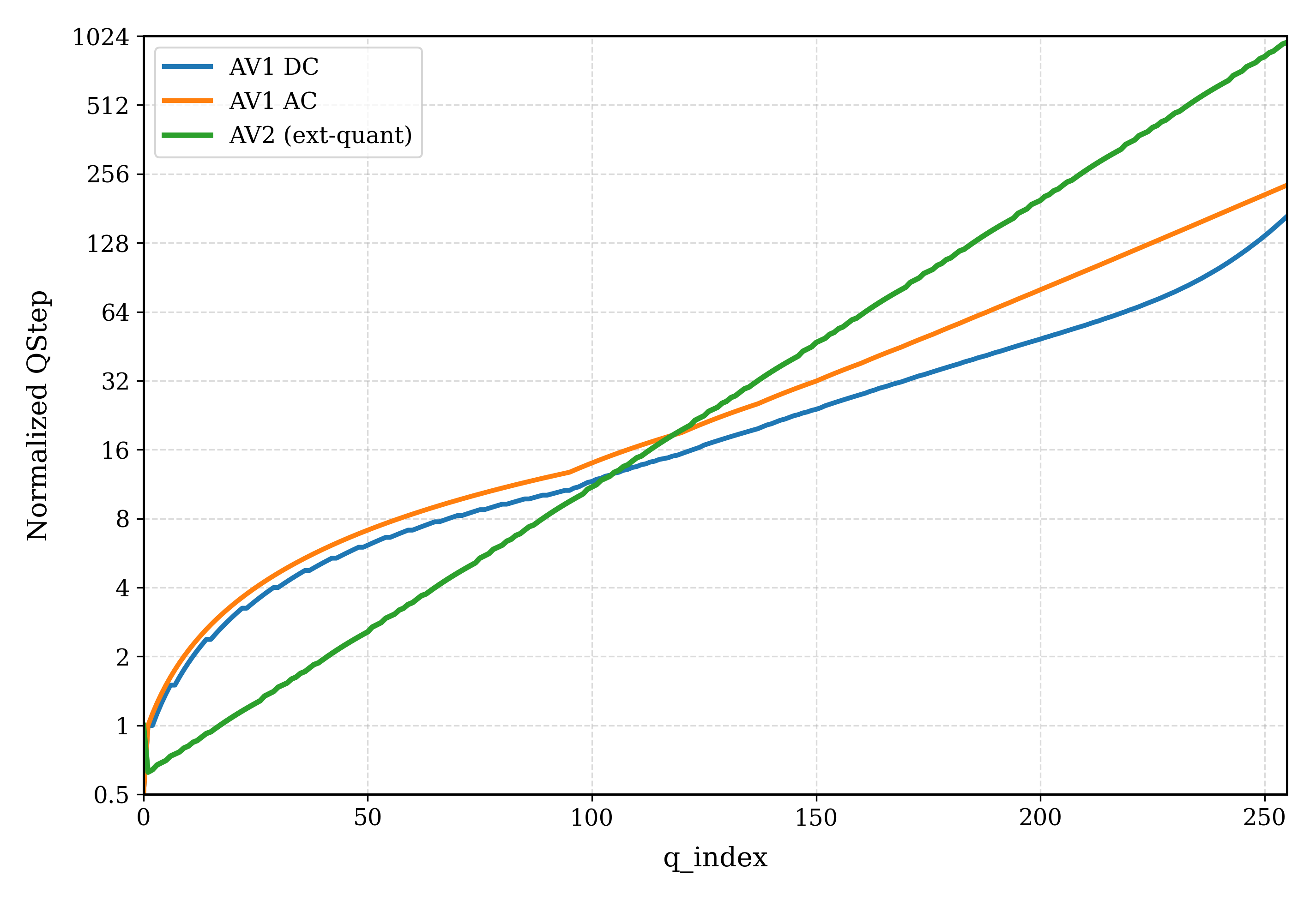}
\caption{Mapping between $q_{\text{index}}$ and normalized QStep for AV1 (AC/DC) and AV2 extended quantization. AV1 QStep values are
normalized by 8, while AV2 QStep values are normalized by 64 to account for the quantization precision.}
\label{fig:av1_av2_qstep_mapping}
\end{figure}
\subsection{Quantization Design in AV1}\label{av1:quantization}
AV1 inherits its quantization design largely from VP9.
In the frame header, a \texttt{base\_q\_idx} is encoded to specify the base frame quantization index,
which is used for luma AC coefficients and also as the base value for other quantizers.
The valid data range for \texttt{base\_q\_idx} is $[0, 255]$.

On top of \texttt{base\_q\_idx}, delta values can be coded in the frame header to
specify the effective quantization index for other quantizers, including DC
coefficients and chroma planes.
The final quantization index $q_{\text{index}}$ for other quantizers is calculated as:
\begin{equation}
q_{\text{index}} = \text{clip}(0, 255, \texttt{base\_q\_idx} + \delta)
\end{equation}

The resulting $q_{\text{index}}$ is an index into a lookup table that defines the
actual quantization step size (QStep).
In AV1, for different internal bit depths (i.e., 8/10/12 bits), six different quantization step size lookup tables (\texttt{Dc\_Qlookup} and \texttt{Ac\_Qlookup}) are defined separately for DC and
AC coefficients.

Figure~\ref{fig:av1_av2_qstep_mapping} illustrates the mapping between
$q_{\text{index}}$ and the QStep value for AC and DC coefficients of 8-bit video
in AV1, where QStep is shown on a logarithmic scale.
The inverse quantization process is formulated as:
\begin{equation}
\text{Denom} =
\begin{cases}
2 & \begin{aligned}[t]
    &\text{if } \texttt{tx\_size} \in \{32\times32, 16\times32, 32\times16, \\
    &\phantom{\text{if } \texttt{tx\_size} \in \{} 16\times64, 64\times16\}
    \end{aligned} \\
4 & \text{if } \texttt{tx\_size} \in \{64\times64, 32\times64, 64\times32\} \\
1 & \text{otherwise}
\end{cases}
\end{equation}
\begin{equation}
\text{coeff} =
\frac{\text{sign} \times ((\text{level} \times \text{QStep}) ~\&~
\texttt{0xFFFFFF})}{\text{Denom}}
\end{equation}
\begin{equation}
\text{coeff} =
\text{clip}\!\left(-2^{7+\text{BitDepth}},
2^{7+\text{BitDepth}} - 1,\,
\text{coeff}\right)
\end{equation}
where \texttt{level} denotes the quantized absolute coefficient magnitude, and
the denominator depends on the transform block size.

\subsection{Entropy Coding Engine}
AV1 uses a context-based multi-symbol arithmetic coder (MS-AC) to efficiently encode syntax elements in the bitstream \cite{han2021av1}.
Unlike binary arithmetic coders (BAC) which can represent only a single binary outcome per coding step \cite{langdon1984introduction,said2003arithmetic}, the MS-AC engine in AV1 supports up to 16 symbols per syntax element with probabilities represented using cumulative distribution function (CDF) tables.
MS-AC improves coding efficiency for syntax elements with incremental values by avoiding the explicit binarization traditionally required in BACs.

In AV1, each syntax element is associated with a context model represented as a CDF table stored with 15-bit precision which are updated adaptively during decoding based on past symbol statistics across the bitstream.

The MS-AC engine is applied to a wide range of syntax elements, including coefficient magnitudes (levels), motion vector components, prediction mode indices, transform type syntax and nearly all syntax elements produced by the codec.
For certain syntax elements, bypass coding is used to further reduce memory requirements and improve throughput, which assumes a uniform probability model without adaptation and dependency to a CDF table.

\subsection{Coefficient Level Map and Sign Coding}\label{av1:level_map}
The transform compacts residuals into a smaller set of transform coefficients. These coefficients are then quantized and coded with a \emph{level-map} representation, where the \emph{level} denotes coefficient magnitude.

The coefficient coding process in AV1 begins by checking whether the TB contains any nonzero coefficients.
If all coefficients are quantized to zero, an \texttt{all\_zero} syntax element is signaled with a value of `1' as the first element of the TB-level syntax which terminates future coding of coefficient related syntax.
If the block contains at least one nonzero coefficient \texttt{all\_zero} sends a value of `0', and an end-of-block (EOB) syntax element is coded in the bitstream.
The EOB syntax is transmitted using a hierarchical scheme that identifies a coarse group of coefficient positions followed by finer bins to determine the exact location of the last significant coefficient. The number of bits required for EOB signaling depends on residual energy compaction with fewer bits spent when coefficients are clustered close towards the DC term.

\begin{figure}[!t]
\centering
\includegraphics[width=0.90\linewidth]{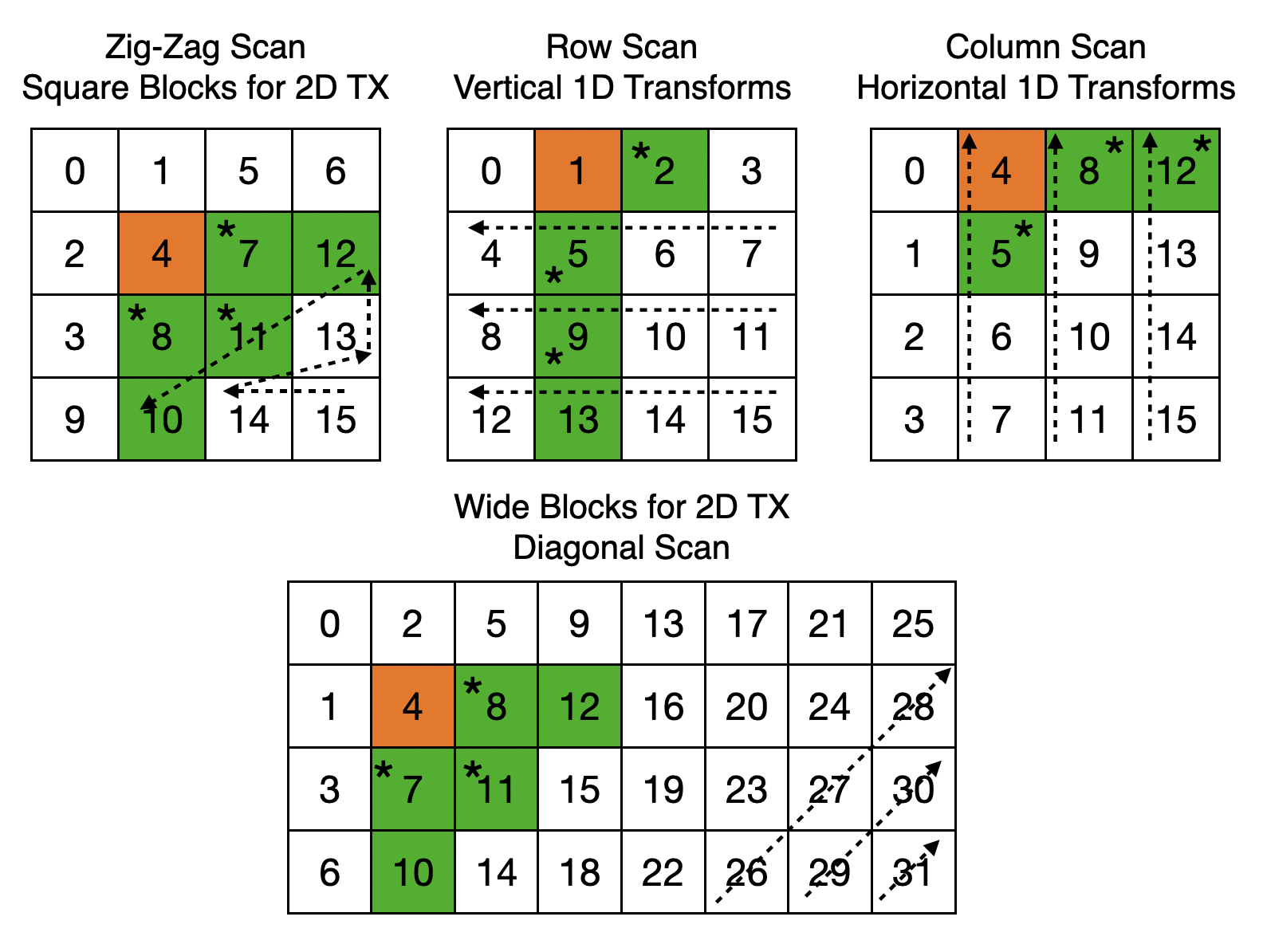}
\caption{AV1 scan orders and context neighborhoods (green) for coefficient coding; numbers show forward scan rank, and the orange cell is the current coefficient to be coded.}
\label{fig:av1_coeff}
\end{figure}

Once the EOB syntax is coded, transform coefficients are signaled in a pre-determined scan order.
AV1 uses several coefficient scan orders depending on transform type and size: zig-zag, up-right diagonal, bottom-left diagonal, row, or column.
One-dimensional transform types (horizontal or vertical) use row or column scans, while two-dimensional transforms use diagonal or zig-zag orders, as illustrated in Fig.~\ref{fig:av1_coeff}.

The absolute magnitudes of coefficients (i.e. levels) are coded starting from the EOB position and proceeding backwards toward the DC term. This framework codes coefficients in three different passes:

\begin{itemize}
    \item \textit{Base Range (BR):} A 4-symbol syntax element that is context-coded using neighboring decoded levels (Fig.~\ref{fig:av1_coeff}).
    This range covers level values $\{0, 1, 2\}$ with an additional escape symbol indicating magnitudes $\geq 3$ which determines whether an additional pass is needed.
    This pass uses a reverse scan order.
    \item \textit{Low Range (LR):} A 4-symbol syntax element that may iterate up to four times, extending the representation to levels $[3,\ldots,14]$ with an escape for level $\ge 15$.
    LR uses a reduced context neighborhood (asterisks in Fig.~\ref{fig:av1_coeff}) and reverse scan order.
    \item \textit{High Range (HR):} For magnitudes $\geq 15$, the residual magnitude beyond 15 is bypass-coded using exponential-Golomb codes. This pass uses a forward scan order different than the BR and LR passes.
\end{itemize}
Coefficient signs are bypass-coded in the HR stage, except that the DC coefficient uses a dedicated sign context.

\section{AV2 Transform and Entropy Coding Design}\label{sec:av2-overview}
This section provides a detailed introduction of the new coding tools and transform-related improvements in AV2.
\begin{figure}[!b]
  \centering
  \includegraphics[width=\columnwidth]{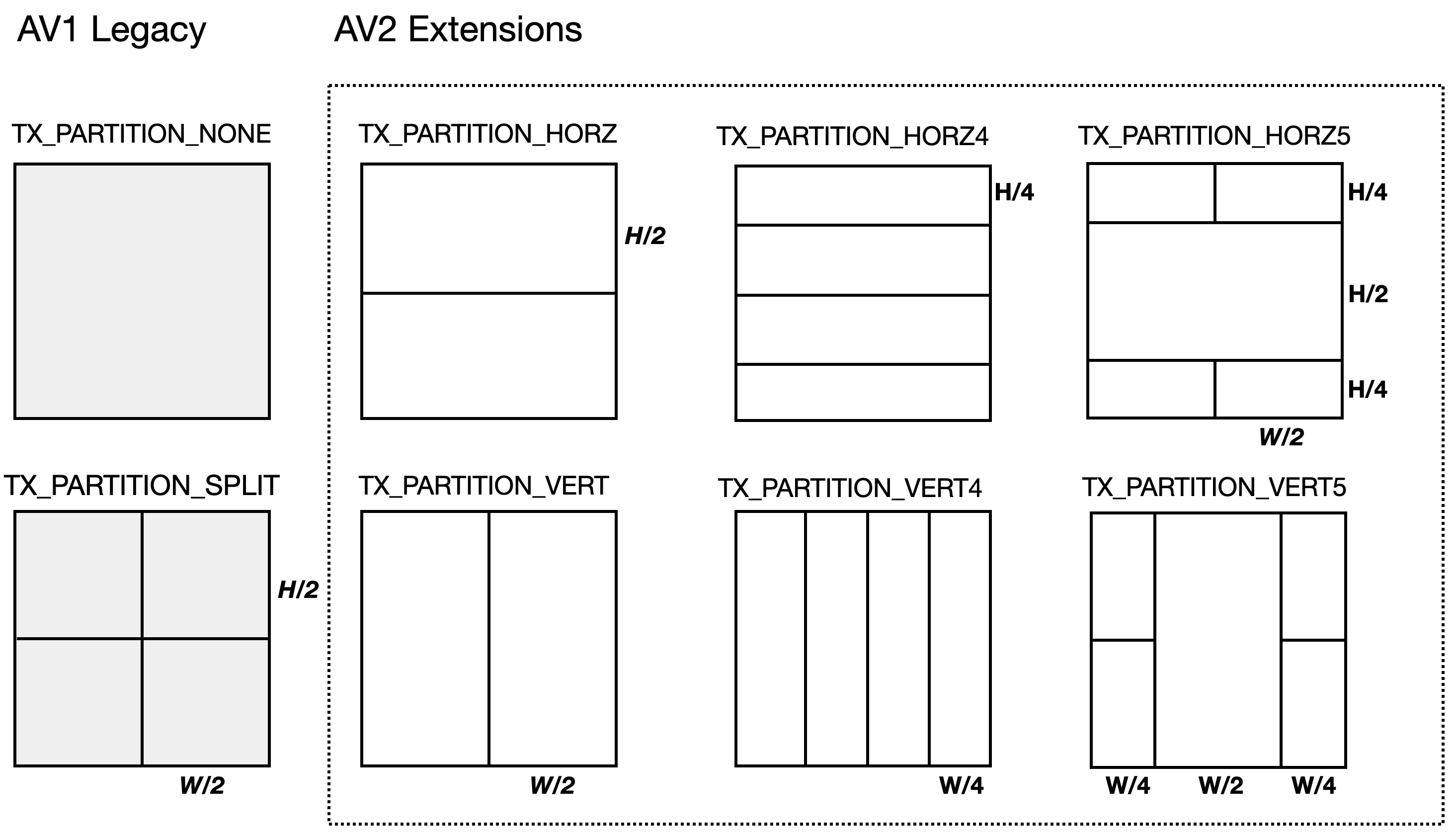}
  \caption{Legacy AV1 transform partition types (\texttt{NONE}, \texttt{SPLIT}) and AV2 extensions (\texttt{HORZ}, \texttt{VERT}), (\texttt{HORZ4}, \texttt{VERT4}) and (\texttt{HORZ5}, \texttt{VERT5}).}
\label{fig:tx-partitions}
\end{figure}

\subsection{Transform Partitioning Improvements}\label{sec:tx-partitioning}
AV2 removes the recursive quad-tree transform partitioning scheme in AV1 which significantly reduces memory usage.
A new partition scheme is introduced which adds six additional partition types over AV1 that applies to both intra and inter blocks in a unified fashion.

In total, the AV2 partitioning design supports eight transform partition types in total as illustrated in Figure \ref{fig:tx-partitions} and with partition types:
\begin{itemize}
  \item \textbf{\texttt{NONE}}: one TB of size $W \times H$ (TB size $=$ CB size)
  \item \textbf{\texttt{SPLIT}}: four TBs of size $W/2 \times H/2$
  \item \textbf{\texttt{HORZ}}: two TBs of size $W \times H/2$
  \item \textbf{\texttt{VERT}}: two TBs of size $W/2 \times H$
  \item \textbf{\texttt{HORZ4}}: four TBs of size $W \times H/4$
  \item \textbf{\texttt{VERT4}}: four TBs of size $W/4 \times H$
  \item \textbf{\texttt{HORZ5}}: rotated H shaped layout with five TBs:
        Top row: two TBs, each $W/2 \times H/4$,
        Middle row: one TB of size $W \times H/2$,
        Bottom row: two TBs, each $W/2 \times H/4$.
  \item \textbf{\texttt{VERT5}}: H shaped layout with five TBs:
        Left column: two TBs, each $W/4 \times H/2$,
        Center column: one TB of size $W/2 \times H$,
        Right column: two TBs, each $W/4 \times H/2$.
\end{itemize}
The new transform partitioning types do not introduce new transform sizes.
Instead, they offer increased flexibility in TB size selection when applicable, enabling diverse decisions for transform type and coefficient coding.
Furthermore, AV2 supports thinner and wider TB shapes with aspect ratios of $1{:}8$ and $1{:}16$ compared to AV1.

\paragraph*{Partition Type Signaling}
Three syntax elements are context coded to signal the transform partition type to the decoder.
First, both the encoder and decoder perform CB size checks determine if horizontal and/or vertical transform splits are allowed.
If a split is allowed in at least one direction, a binary symbol, \texttt{do\_partition} is coded.
This indicates whether the CB has a TB split.
If \texttt{do\_partition} indicates no partitioning, the TX partition type is inferred as \texttt{NONE}.

If \texttt{do\_partition} indicates a split and both horizontal and vertical directions are allowed for the given block size, a 7 symbol syntax element, \texttt{txfm\_4way\_partition\_type}, selects one of the 7 transform partition types illustrated in Fig. \ref{fig:tx-partitions} excluding NONE.
Otherwise, if only a single direction is allowed (e.g., only vertical or only horizontal), \texttt{txfm\_2or3\_way\_partition\_type} is coded.
This signals whether the quarter strip along the permissible axis is used, for example, choosing between \texttt{HORZ} and \texttt{HORZ4}.

\begin{table}[!b]
\centering
\caption{Properties of Primary Transform Types in AV2}
\label{tab:av2_transform_types}
\renewcommand{\arraystretch}{1.05}
\setlength{\tabcolsep}{3pt}
\scriptsize
\begin{tabular}{|c|c|c|c|}
\hline
\textbf{TX Type} & \textbf{Size (N-point)} & \textbf{Operation} & \textbf{Fwd/Inv Bit Width} \\
\hline
DCT-2     & 4/8/16/32 & BTF / MatrixMult & 8-bit  \\
DST-4     & 4            & BTF / MatrixMult & 8-bit  \\
L-ADST    & 8            & MatrixMult       & 8-bit  \\
DST-7     & 16           & MatrixMult       & 8-bit  \\
DDT       & 8/16       & MatrixMult       & 8-bit  \\
IDTX      & 4/8/16/32    & Single Multiply  & 8-bit  \\
\hline
\end{tabular}
\end{table}

\subsection{Primary Transforms and AV2 Improvements}\label{sec:primary-transforms}
AV2 retains the same set of 16 separable primary transform combinations as AV1, as summarized in
Table~\ref{tab:tx-types-av2}, while introducing several refinements in transform kernel design and
transform type signaling. The main improvements include:

\begin{itemize}
	\item Redesign of core transform kernels using 8-bit multipliers with matrix-based arithmetic,
    while allowing butterfly-based factorizations as optional implementations for selected kernels
    (e.g., DCT-2).
	\item Removal of the native 64-point primary transform and replacement with a 32-point
    transform combined with residual upsampling for $64\times N$ and $N\times 64$ transform blocks.
	\item Extension of $64\times N$ and $N\times 64$ transform blocks to chroma residual coding for
    4:2:0, 4:2:2, and 4:4:4 formats.
    \item Revision of the ADST family, replacing the 4-/8-/16-point ADSTs with DST-4, a learned
    unitary ADST (L-ADST), and DST-7 kernels to improve coding efficiency.
	\item Introduction of data-driven transforms (DDTs) for inter-predicted blocks.
    \item Unified and context-adaptive transform type signaling conditioned on block size and
    prediction mode.
\end{itemize}

\paragraph*{AV2 Transform Improvements}
In AV2, primary transform kernels are implemented using fixed integer arithmetic with 8-bit
multipliers and 16-bit accumulators, as summarized in Table~\ref{tab:av2_transform_types}. Compared to AV1, this design removes stage-dependent rules and simplifies control logic.

For transform block (TB) sizes of 8 and above, the DCT-2 is implemented using fixed integer
matrices, enabling parallel computation and avoiding multi-stage processing. For the 4-point
case, compact butterfly-based implementations may still be used for DCT-2 and DST-4 without
affecting the bitstream.

To improve coding efficiency, the 8-point and 16-point ADSTs used in AV1 are replaced by a
learned unitary ADST (L-ADST) and a fixed DST-7 kernel, respectively. These kernels provide
improved energy compaction while maintaining low arithmetic precision.

\paragraph*{Largest Transform with Upsampling}
A notable feature of the AV2 transform design is the removal of the native 64-point primary
transform. For $64\times N$ and $N\times 64$ transform blocks, AV2 applies a 32-point inverse
transform followed by a residual upsampling process implemented via sample duplication along
the long dimension. This approach achieves coding performance comparable to a full 64-point
transform while substantially reducing decoder complexity.

The same 32-point transform combined with residual upsampling is applied to chroma residual
coding, enabling the use of large transform blocks for chroma without introducing native
64-point transform logic. Experimental results show that enabling large chroma transform blocks
yields significant coding gains at high resolutions.

\paragraph*{Inter Data-Driven Transforms (DDTs)}
AV2 introduces data-driven transforms (DDTs) for $8\times8$ and $16\times16$ inter blocks.
These kernels are trained offline via Karhunen–Loève Transformation (KLT) on prediction residuals.
Each DDT includes a learned kernel and its flipped variant (e.g., DDT8, FLIPDDT8).
DDTs use the same transform type syntax as FLIPADST/ADST and integrate cleanly into the transform pipeline, e.g., they are used in-place of ADST variants shown in Table \ref{tab:tx-types-av2} reusing the same syntax.
For inter blocks, the 8-point DDT replaces the learned ADST (L-ADST), and the 16-point DDT replaces the fixed DST-7 kernel.
Directional transforms DST-7 and L-ADST still remain available for intra-coded blocks.

\begin{figure}[!b]
  \centering
  \includegraphics[width=0.7\columnwidth]{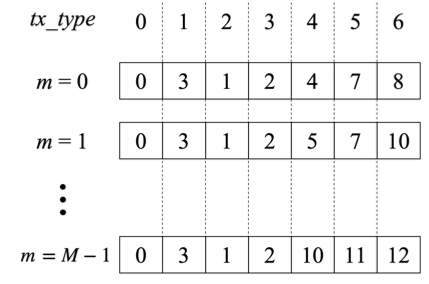}
  \caption{ An example list of transform identifiers depending on the set index, which shows the mapping from signaling index ($\textit{tx\_type}$) to transform candidates for different $m$, where $M$=39 in the AV2 design. The transform candidates $0$ and $3$ denote \texttt{DCT\_DCT} and \texttt{ADST\_ADST}, respectively.}
\label{fig:ITSD-table}
\end{figure}

\subsection{Intra Mode Dependent Transform Set Derivation (MDTX)}\label{sec:tx_signaling}
For intra blocks, AV2 restricts transform type signaling to a subset of transform types for each TB to reduce bitstream overhead.
A mode dependent transform set derivation (MDTX) framework is introduced as shown in Figure \ref{fig:ITSD-table}, where a mode $m$ determines the particular transform set index containing 7 candidate transform types, and numbers in each set correspond to transform id's previously shown in Table \ref{tab:tx-types-av2}.

This approach generalizes the transform set definitions of AV1 shown in Table \ref{tab:av1-tx-sets-a} for intra coded TBs and allows for determining the optimal set based on empirical pre-trained data.

Every transform set always includes the two baseline transforms \texttt{DCT\_DCT} and \texttt{ADST\_ADST}.
The remaining five transform types in each set are derived based on the intra prediction mode and block size according to:
\begin{itemize}
\item 9 angular intra prediction modes: \texttt{DC\_PRED}, \texttt{V\_PRED}, \texttt{H\_PRED}, \texttt{D45\_PRED}, \texttt{D135\_PRED}, \texttt{D113\_PRED}, \texttt{D157\_PRED}, \texttt{D203\_PRED}, \texttt{D67\_PRED}, and 4 smooth and paeth intra prediction modes: \texttt{SMOOTH\_PRED}, \texttt{SMOOTH\_V\_PRED}, \texttt{SMOOTH\_H\_PRED}, and \texttt{PAETH\_PRED}.
\item 3 block-size groups based on the TB width (w) and height (h), consistent with the TB size groups of AV1 as in Table~\ref{tab:av1-tx-sets-b}:
  \begin{itemize}
  \item min(w, h) $=$ 4 and max(w, h) $<$ 32
  \item min(w, h) $=$ 8 and max(w, h) $<$ 32
  \item min(w, h) $=$ 16 and max(w, h) $<$ 32
  \end{itemize}
\item As in AV1, only \texttt{DCT\_DCT} is used for block sizes largest blocks that satisfy max(w, h) $\in$ {32, 64}.
\end{itemize}

In total, 39 classes (belonging to 3 block-size groups $\times$ 13 intra modes) are defined for transform set derivation (i.e. $M=39$ in Figure \ref{fig:ITSD-table}).
At the decoder, the decoded $\textit{tx\_type}$ is mapped through a $39\times7$ lookup table, using block size and intra mode (i.e., $m$) to determine the applied transform.

\subsection{Intra/Inter Secondary Transforms}\label{sec:ist}
\begin{figure*}[!t]
\centering
\includegraphics[width=\textwidth,height=\textheight,keepaspectratio]{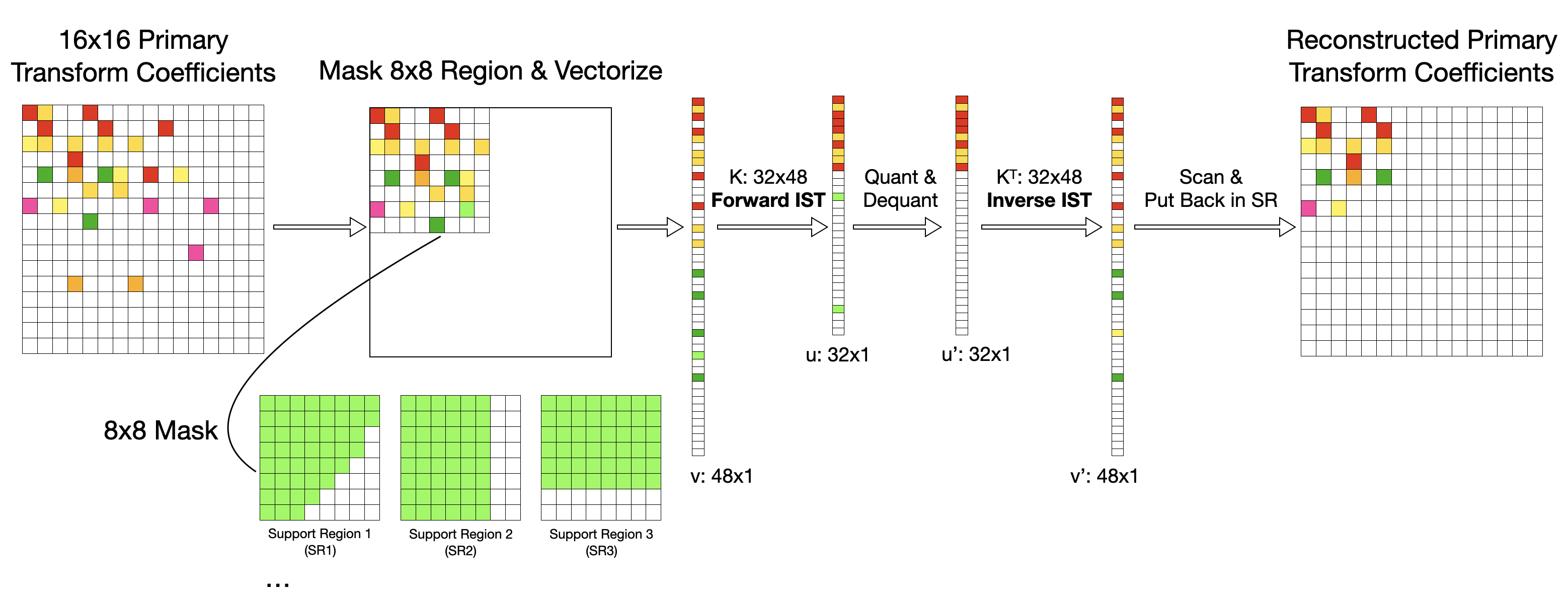}
\caption{Block diagram illustrating the operation of the intra/inter secondary transform (IST) and examples of predefined support regions.}
\label{fig:SR}
\end{figure*}
Separable spatial transforms constitute a fundamental tool for decorrelating prediction residuals in major video coding standards/formats, including H.264/AVC~\cite{Wieg2003}, HEVC~\cite{sull2012}, AV1~\cite{han2021av1}, and VVC~\cite{xin2021}.
However, their effectiveness is limited for residual signals containing arbitrarily oriented edges and complex textures commonly observed in natural video content.
Non-separable transforms, while capable of better modeling such patterns, incur significantly higher computational complexity~\cite{Arru2014,Taka2013}.
To mitigate this cost, non-separable secondary transforms first introduced in~\cite{Xin2016}, are applied as an intermediate stage between the primary transform and quantization.
Further refinements, including restricting the transform to low-frequency coefficients, zeroing out high-frequency components, and limiting kernel sizes~\cite{Koo2019}, have made these approaches practical for modern codecs.
In AV2, Intra/Inter Secondary Transforms (IST) are introduced as part of the AV2 transform coding framework to improve residual energy compaction while maintaining manageable complexity.

The application of forward and inverse IST on a $16\times16$ primary transform block is illustrated in Figure~\ref{fig:SR}.
In the forward IST stage, a predefined support region located in the upper-left corner of the primary transform block is selected.
For transform blocks of size $\geq 8\times8$, this support region corresponds to a 64-coefficient area, while for transform blocks smaller than $8\times8$, a reduced 16-coefficient region is used.
The exact shape of the support region is kernel-dependent and predefined; several representative examples are shown in Figure~\ref{fig:SR}.

Primary transform coefficients within the selected support region are masked and vectorized to form an input vector
$\mathbf{v} \in \mathbb{R}^{N}$, where $N$ depends on the support region and transform block size.
As illustrated in Figure~\ref{fig:SR}, for a $16\times16$ block using DCT-2 as the primary transform, the support region produces a vector of length $N=48$.
This vector is processed by the forward intra/inter secondary transform using a kernel matrix
$\mathbf{K} \in \mathbb{R}^{M \times N}$, yielding
\begin{equation}
\mathbf{u} = \mathbf{K}\mathbf{v},
\end{equation}
where $\mathbf{u} \in \mathbb{R}^{M}$ denotes the intra/inter secondary transform coefficients.
Only a reduced number of basis vectors is retained ($M < N$), while higher-frequency components are implicitly set to zero, consistent with the dimensionality reduction shown in Figure~\ref{fig:SR}.

The intra/inter secondary transform coefficients $\mathbf{u}$ are subsequently quantized and entropy coded as described in the following sections.
At the decoder, the parsed and dequantized coefficients $\hat{\mathbf{u}}$ are processed by the inverse intra/inter secondary transform using the transpose of the same kernel:
\begin{equation}
\hat{\mathbf{v}} = \mathbf{K}^{T}\hat{\mathbf{u}}.
\end{equation}
The reconstructed vector $\hat{\mathbf{v}}$ is then scanned and placed back into the corresponding support region locations within the primary transform block, while coefficients outside the support region remain unchanged.
This process reconstructs the primary transform coefficient block prior to the inverse primary transform stage.

\input{tables/atc_table4_st}

Some key implementation details are summarized as follows:
\begin{itemize}
    \item IST is applied to a) luma intra blocks when both horizontal and vertical primary transforms are either DCT-2 or ADST; b) luma inter blocks when both horizontal and vertical primary transforms are DCT-2.
    \item Reduced basis representations are used: $8\times16$ for transform blocks smaller than $8\times8$, $32\times48$ for transform blocks $\geq 8\times8$ using DCT-2, and $20\times48$ for transform blocks $\geq 8\times8$ using ADST. Higher-frequency components are zeroed.
    \item Multiple kernel sets are defined, each containing a number of kernels. Both the set index and kernel index are explicitly signaled, decoupling kernel selection from prediction mode signaling.
    \item For luma inter blocks, a fixed kernel set (set index $=0$) is used, while the kernel index remains signaled.
\end{itemize}

Table~\ref{tab:st_memory} summarizes the ROM memory requirements for storing the IST kernels, along with the corresponding worst-case computational complexity expressed in multiplications per pixel.

\subsection{Cross-chroma Component Transforms (CCTX)}\label{sec:cctx}
The well-known YCbCr color space provides good perceptual uniformity and overall compression efficiency.
However, local correlations can still exist among the three YCbCr planes. While existing tools such as chroma from luma (CfL) \cite{trudeau2018predicting} effectively reduce the luma-chroma correlation, a new tool called cross-chroma component transforms (CCTX) has been introduced in AV2 to specifically address the correlation between the two chroma planes (CbCr) after a primary transforms.

Given a pair of colocated chroma transform coefficients ($x_u$, $x_v$) at a particular spatial position, a 2D rotation is applied:
\begin{equation}
\label{eq:cctx_matrix}
\begin{pmatrix} x_{C1}\\ x_{C2} \end{pmatrix} = \begin{pmatrix} \cos\theta & \sin\theta \\ -\sin\theta  & \cos\theta \end{pmatrix} \begin{pmatrix} x_{u}\\ x_{v} \end{pmatrix}.
\end{equation}
For the choice of rotation angle $\theta$, seven angles are used: $0\degree$, $45\degree$, $30\degree$, $60\degree$, $-45\degree$, $-30\degree$, and $-60\degree$.
The corresponding transform operations are converted to integer space, with the matrix elements from Equation \eqref{eq:cctx_matrix} implemented with 9-bit precision.

The CCTX mode is signaled in the bitstream jointly to represent both chroma channels at the TB level.
The resulting coefficients $x_{c1}$ and $x_{c2}$ replace the $x_u$ and $x_v$ coefficients before moving to the coefficient coding stage.
Notably, when $\theta=0$, the rotation is an identify transform, and the $x_u$ and $x_v$ remain unchanged.

When two rotation angles differ by $90\degree$, their resulting coefficient pairs, C1 and C2, have the same magnitudes but are swapped between the two planes. For instance, the C1 coefficient from a $30\degree$ rotation has the same magnitude as the C2 coefficient from a $-60\degree$ rotation, with a sign flip.
This property reduces signaling overhead since there cannot be an all-zero C1 plane and a non-zero C2 plane.

\subsection{DC-based Transform Restriction (DCTX)}\label{sec:dctx}
To reduce signaling overhead associated with transform-type signaling and to improve compression efficiency, AV2 introduces a DC-based transform signaling restriction intra blocks.

If a TB contains only a DC coefficient (i.e., all other coefficients are quantized to zero except for the 0th coefficient), then AV2 skips both primary transform type and intra/inter secondary transform (IST) related signaling.
The last position or end-of-block (EOB) syntax signaling is moved ahead of transform type signaling in AV2 compared to AV1 design.

For chroma channels where only the DC coefficient is non-zero, AV2 further prevents signaling of syntax associated CCTX.
This reduces signaling overhead for smooth chroma regions and improves compression efficiency in high-resolution and screen content scenarios.

\subsection{Quantization Design in AV2}\label{sec:av2_quantization}
As described in Section~\ref{av1:quantization}, AV1 employs lookup tables to define
the \texttt{q\_index} to QStep mapping for different bit depths and coefficient
types (AC/DC). This results in a non-trivial mapping between QStep and
\texttt{q\_index}, as shown in Figure~\ref{fig:av1_av2_qstep_mapping}.

AV2 replaces this logic with a closed-form exponential formulation, providing a
clear mathematical relationship between \texttt{q\_index} and QStep while reducing
memory requirements in both encoder and decoder implementations, as also shown in
Figure~\ref{fig:av1_av2_qstep_mapping}. This makes the relationship between
\texttt{q\_index} and bitrate more predictable across quantization levels,
simplifying rate-control logic for different applications.

AV2 also extends the maximum quantization step size beyond the range supported by
AV1, as illustrated in Figure~\ref{fig:av1_av2_qstep_mapping}. This enables
low-latency applications that require aggressive bitrate reductions in
bandwidth-constrained scenarios.

\paragraph{Unified equation for \texttt{q\_index} to QStep mapping}
AV2 replaces the six separate QStep lookup tables for AC and DC coefficients in
AV1 with a single unified exponential function. For 8-bit video, the following
piecewise equation defines the mapping between \texttt{q\_index} and QStep:
\begin{equation}
\text{QStep}(q) =
\begin{cases}
32 & q = 0 \\[0.5ex]
\text{round}\!\left(2^{(q + 127)/24}\right) & q \in [1, 24] \\[0.5ex]
\begin{aligned}[t]
&[\text{QStep}((q-1) \bmod 24) + 1] \\
&\quad \times 2^{\lfloor (q-1)/24 \rfloor}
\end{aligned} & q \geq 25
\end{cases}
\end{equation}

This formulation ensures that the quantization step size doubles whenever
\texttt{q\_index} is incremented by 24, providing consistent bitrate behavior
across the valid \texttt{q\_index} range. The QStep value at $q = 0$ is set to 32
to preserve the lossless coding behavior inherited from AV1, accounting for the
gain of 4 from the $4\times4$ Hadamard transform used in lossless mode.

To prevent the exponential formulation from producing duplicate QStep values at
low \texttt{q\_index} ranges, the QStep precision is increased by $8\times$
compared to AV1.

\paragraph{Delta \texttt{q\_index} Offset for DC Coefficients}
AV2 additionally allows signaling delta offsets at the sequence level to support
independent quantization of DC and AC coefficients. Specifically, a 5-bit
\texttt{base\_y\_dc\_delta\_q} value and a 5-bit
\texttt{base\_uv\_dc\_delta\_q} value in the range $[-8, 23]$ are coded to indicate
the offset of the \texttt{q\_index} for luma and chroma DC coefficients relative to
AC coefficients:
\begin{equation}
\begin{aligned}
\texttt{q\_idx}_{\text{Y\_DC}} &= \text{clip}(0, 255, \texttt{base\_q\_idx} \\
&\quad + \texttt{base\_y\_dc\_delta\_q}) \\
\texttt{q\_idx}_{\text{UV\_DC}} &= \text{clip}(0, 255, \texttt{base\_q\_idx} \\
&\quad + \texttt{base\_uv\_dc\_delta\_q})
\end{aligned}
\end{equation}
This provides independent DC/AC quantization control for luma and chroma,
improving visual quality in smooth regions without requiring separate
quantization tables.

\paragraph{Quantization for High Bit-Depth Video}
For higher bit depths, including 10-bit and 12-bit video, AV2 adopts an
offset-based approach to match bitrates across different bit depths while
enabling access to the smallest QStep values at higher precision. This introduces
two normative changes:
\begin{itemize}
    \item The \texttt{q\_index} syntax is extended from 8 bits to 9 bits for
    10-bit and 12-bit operation. The operational ranges become $[0,255]$ for
    8-bit, $[0,303]$ for 10-bit, and $[0,351]$ for 12-bit.
    \item When \texttt{q\_index} exceeds 255, the value is offset by $-48$ for
    10-bit and $-96$ for 12-bit, and the corresponding QStep is scaled by factors
    of 4 and 16, respectively.
\end{itemize}
At the encoder, to match bitrates across 8/10/12-bit operation, an offset is added
to \texttt{q\_index}, and the modified \texttt{q\_index} is signaled in the
bitstream.

\subsection{Trellis Coded Quantization}\label{sec:tcq}

Trellis coded quantization (TCQ) is a well-studied advanced quantization method that can improve coding efficiency with a modest increase in complexity~\cite{Marcellin1990TCQ}.
%
AV2 supports flexible signaling of quantization methods -- between the default scalar quantization (SQ), TCQ, and Parity Hiding (PH) methods -- at both the sequence and frame levels.
When TCQ is enabled for a frame, it is applied to luma transform blocks (TBs) with 2D scan, whereas TBs with 1D scan, chroma blocks, and blocks with Forward Skip Coding (FSC) all fall back to SQ.

\begin{figure}[!t]
\centering
\includegraphics[width=0.48\textwidth]{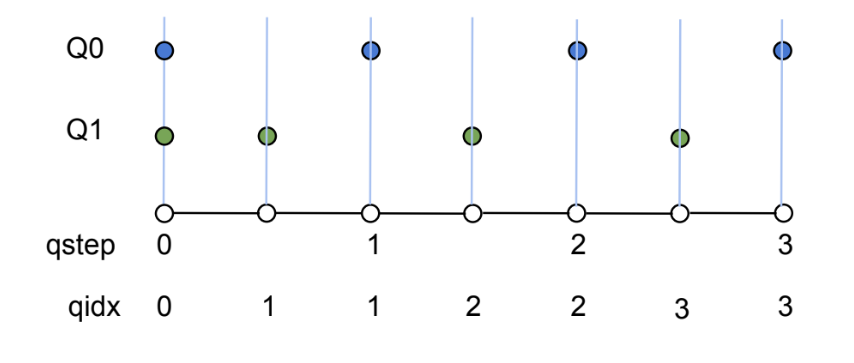}
\caption{TCQ with dual quantizers Q0 and Q1}
\label{fig:tcq_quant}
\end{figure}

\begin{figure}[!b]
\centering
\includegraphics[width=0.48\textwidth]{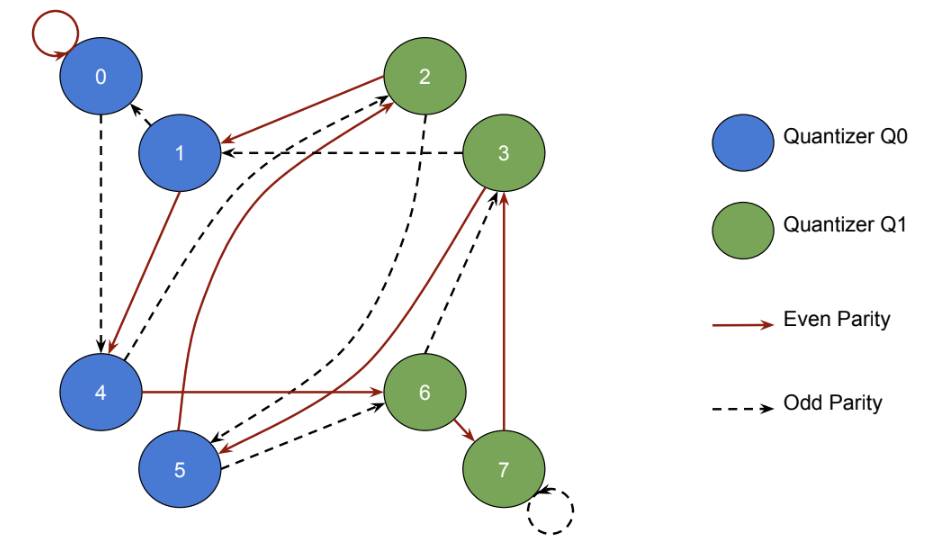}
\caption{TCQ state machine.}
\label{fig:tcq_fsm}
\end{figure}

Similar to existing methods -- such as the version adopted by JPEG 2000 extensions~\cite{t.801} -- the TCQ mode in AV2 defines two underlying sub-quantizers, labeled as Q0 and Q1.
As Fig.~\ref{fig:tcq_quant} illustrates, the non-zero reconstruction levels of Q1 are offset by a one-half step relative to Q0.
The sub-quantizer selection is governed by a finite-state machine where each state is associated with either Q0 or Q1 (see Fig.~\ref{fig:tcq_fsm}).

The initial state is reset to zero at the start of decoding each TB. As coefficients are processed in scan order, their magnitudes and parities are decoded using novel syntax that ensures consistent parity across multiple syntax passes. The parity bit of the decoded coefficient (i.e., even or odd) dictates the transition to one of two possible next states, which in turn determines which sub-quantizer will be applied for the next coefficient.


\begin{figure}[!t]
\centering
\includegraphics[width=0.48\textwidth]{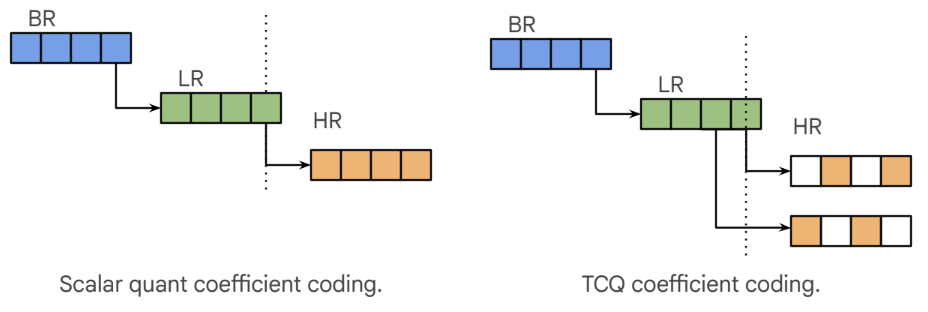}
\caption{Low (BR), mid (LR) and high (HR) syntax for scalar and TCQ coding.}
\label{fig:tcq_syntax}
\end{figure}

\subsubsection{Syntax Adjustments for Coefficient Coding}
TCQ mode requires minor adjustments to the coefficient coding syntax.
AV2 utilizes distinct syntax ranges for coefficient values: a base range (BR), a low-level range (LR), and a high range (HR). The highest value in both the BR and LR syntax serves as an escape code, signaling the use of the subsequent syntax level.
Coefficient syntax is transmitted in two passes over the coefficient block. The first pass conveys the BR and LR syntax, while the second pass is dedicated to the HR syntax and sign coding. To maintain parity across these passes in TCQ mode, the two highest values of the combined BR and LR ranges are designated as escape codes. This necessitates that the HR syntax be scaled by a factor of two. In addition, TCQ mode uses separate BR probability models corresponding to the Q0 and Q1 quantizers.

\begin{figure}[!b]
\centering
\includegraphics[width=0.48\textwidth]{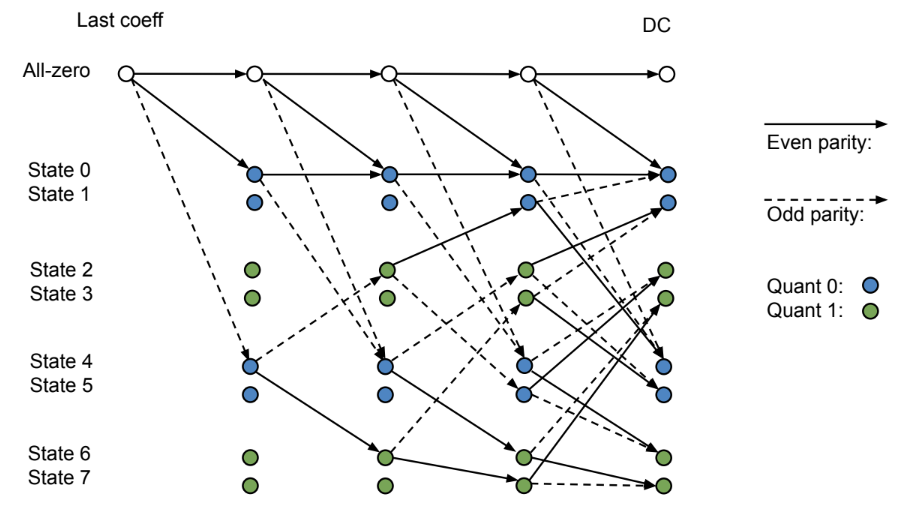}
\caption{TCQ encoder algorithm.}
\label{fig:tcq_alg}
\end{figure}

\subsubsection{Encoder Algorithm}
The reference encoder performs coefficient optimization using the Viterbi dynamic programming algorithm, which jointly optimizes both the quantization decisions and the end-of-block (EOB) position.
The encoder maintains nine states: one for all-zero coefficients, and eight coded states corresponding the states defined in the TCQ state diagram in Figure ~\ref{fig:tcq_fsm}. The algorithm is executed in two passes.

In the first pass, the algorithm proceeds from the last non-zero coded coefficient (i.e. identified through simple quantization) and iterates through each coefficient position in reverse diagonal scanning order.
At each position, quantization decisions (represented by the transition arrows in Figure ~\ref{fig:tcq_alg}) are evaluated to update the rate-distortion costs for each potential next-state.
When multiple decisions lead to the same next-state, only the decision with the lowest rate-distortion cost is retained.
The iterative process continues until the final coefficient position is reached.

After the first pass completes, the final state with the lowest overall rate-distortion cost is determined, and a second backtracking pass is made through the trellis to determine the quantization decisions and EOB position that led to the best result.

\subsubsection{Encoder Runtime Reduction}
The Viterbi algorithm used in the encoder, while efficient, can still lead to a significant increase in encoder runtime.
The AV2 reference encoder employs the following techniques to mitigate this complexity.

First, Single Instruction Multiple Data (SIMD) optimizations are leveraged to parallelize the trellis state evaluations. This approach yields significant reductions in encoding cycles, with a 52\% decrease observed for All-Intra (AI) configurations and a 40\% decrease for Random-Access (RA) configurations. Further vectorization is achieved by processing consecutive coefficients within the same diagonal coordinate in parallel, exploiting the absence of entropy context dependencies between them. Updates across trellis states are also efficiently parallelized with the execution of a small number of SIMD instructions.

Second, the complexity of the trellis search is inherently linked to the number of coefficients evaluated, which is proportional to the initial tentative EOB position. To expedite the encoding process, an EOB pre-truncation technique can be applied. This involves using a simpler, faster search method, such as rate-distortion optimized (RDO) scalar quantization, to determine a reduced EOB position. This pre-truncation effectively limits the search space for the more complex Viterbi algorithm, leading to a reduction in encoding time.

\subsubsection{Quantization Step Size Adjustment}
For a given content and quantization step size QStep, TCQ systematically yields lower quantization distortion than its SQ counterpart. To better accommodate TCQ within the existing rate-distortion optimization (RDO) framework of AV2, the encoder adds an internal offset to \texttt{q\_index} before deriving the value of QStep following the same mapping as in SQ. This offset is empirically chosen as $+2$ for 8-bit sequences and $+4$ for 10-bit sequences, so as to approximately align the expected distortion between SQ and TCQ at the same input \texttt{q\_index}.

\subsection{Probability Adaptation Rate Adjustment (PARA)}\label{sec:para}
\begin{figure}[!t]
\centering
\includegraphics[width=0.44\textwidth]{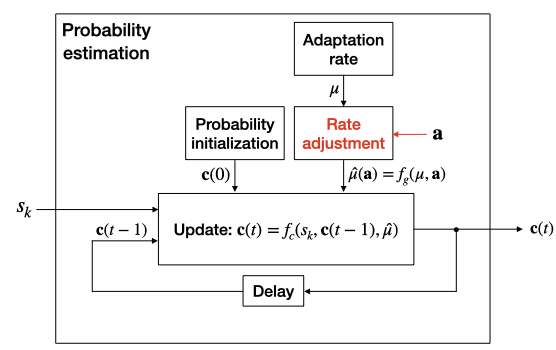}
\caption{The building blocks of the probability estimation steps where the rate adaptation step in red is added to AV2 based on the probability adaptation rate adjustment (PARA) approach.}
\label{fig:entropy_probest_new}
\end{figure}

AV2 reuses the same arithmetic coding engine from AV1 with a novel probability adaptation rate adjustment (PARA) step as shown in Figure \ref{fig:entropy_probest_new}. PARA allows better adaptation of the probability updates for syntax elements.

In both AV1 and AV2, the probability estimation is performed for multiple symbols whose probabilities are represented in terms of cumulative distribution functions (CDFs). The CDF values are represented using 15-bit unsigned integers, obtained by scaling the actual probability range [0, 1] by $2^{15}$

In $M$-ary (multi-symbol) arithmetic coding, the probabilities assigned for $M$ symbols can be represented by the vector of CDF values $\mathbf{c}(t)$ at time $t$ as:
\begin{equation}
    \mathbf{c}(t) = \left[  c_0(t), c_1(t), \dots, c_{M-1}(t) \right]
\end{equation}
where $c_i(t)$ denotes the CDF value at $i$-th symbol at time $t$.
In the probability estimation step of AV2, the CDF update formula can be written in terms of CDFs as:
\begin{equation}
c_i(t) = \begin{cases}
c_i(t-1) - \mu \cdot c_i(t-1)  & i < k \\
c_i(t-1) + \mu \cdot (p_{\text{max}}-c_i(t-1))  & i \geq k \\
\end{cases}
\label{eqn:cdf_update}
\end{equation}
where $k$ denotes the index of the coded symbol element (i.e., index of symbol $s_k$ as shown in Figure~\ref{fig:entropy_probest_new}), and $i$ corresponds to the index of the CDF element associated with the $i$-th symbol updated after $s_k$ is coded.
The adaptivity rate $\mu$ in Equation \ref{eqn:cdf_update} depends on the number of symbols in the group ($M$) and the counter value, reflecting the number of coded symbols in the group.

The derivation of the adaptivity rate in AV1 can be written as:
\begin{equation}
\mu(r_C,r_M) = \left({2^{3+r_C(n) + r_M(m) }}\right)^{-1}
\label{eqn:adaptation_rate}
\end{equation}
where $r_C(n)$ and $r_M(m)$ can be formulated as
\begin{equation}
    r_C(n)=
    \begin{cases}
        0 & \text{if } n \leq 15 \\
        1 & \text{if } 15 < n \leq 31 \\
        2 & \text{if } n > 31
    \end{cases}
    \label{eqn:rate_counter}
\end{equation}
\begin{equation}
    r_M(m)=
    \begin{cases}
        1 & \text{if } m = 2 \text{ or } 3\\
        2 & \text{otherwise}
    \end{cases}
        \label{eqn:rate_length}
\end{equation}
such that $r_C(n)$ is a function of $n$ denoting the number of coded symbols in the encoding/decoding process, and $r_M(m)$ is a function of the number of symbols in the group.

Note that in both Equation \ref{eqn:rate_counter} and Equation \ref{eqn:rate_length} larger values of $r_C(n)$ and $r_M(m)$ lead to smaller $\mu$  so that it results in a slower adaptation rate. On the contrary, smaller values of  $r_C(n)$ and $r_M(m)$ lead to faster adaptation rate in the probability estimation.

In AV2, the PARA concept allows to further adjust the rate parameter $\mu$ as illustrated in Figure~\ref{fig:entropy_probest_new}. Specifically, it modifies the adaptivity rate $\mu(r_C,r_M)$ stated in Equation \ref{eqn:adaptation_rate}, by using an adjustment parameter $\alpha$ as follows:
\begin{equation}
    \widehat{\mu}(r_C,r_M,\alpha) = \left({2^{R + \alpha  }}\right)^{-1}
    \label{eqn:mu_term_with_alpha}
\end{equation}
where $R=3+r_C(n) + r_M(m) $, and the parameter $\alpha$ is added to the exponent as an offset. Note that, for an integer value of $\alpha$, the adjusted rate $\widehat{\mu}(r_C,r_M,\alpha)$ can be implemented efficiently using bit-wise operations since it can be stated in terms of powers of two. Specifically, the multiplications in the CDF update Equation \ref{eqn:cdf_update} can be implemented using a right-shift operator ($\gg$), by replacing the multiplication term $\mu \cdot x$  with $x \gg \mu$ as follows:
\begin{equation}
c_i(t) = \begin{cases}
c_i(t-1) - (c_i(t-1)  \gg (R+\alpha)))  & i < k \\
c_i(t-1) + ((2^{15}-c_i(t-1)) \gg (R+\alpha))  & i \geq k \\
\end{cases}
\label{eqn:cdf_update_bitshifts}
\end{equation}
where $R=3+r_C(n) + r_M(m) $ and $p_{max}= 2^{15}$.

In AV2, separate adjustment parameters (denoted as $\mathbf{a}$ in Figure ~\ref{fig:entropy_probest_new}) are jointly optimized per-context (i.e., group of symbols associated with each row of the CDF table), and different time intervals defined by the $r_C(n)$  in Equation \ref{eqn:rate_counter} so that each time the interval $\mathcal{T}$ is defined as:
\begin{equation}
    \mathcal{T} =
    \begin{cases}
T_0  & \text{if $n\leq 15$ } \\
T_1  & \text{if $15 < n \leq 31$ } \\
T_2  & \text{if $n > 31$} \\
\end{cases}
\end{equation}
where $n$ denotes the number of coded symbols, and the thresholds $15$ and $31$ are the same as the ones used in Equation \ref{eqn:rate_counter} for $r_C(n)$.

In AV2, three separate PARA parameters are assigned for each time interval per context entry (i.e., per symbol group), denoted by $(\alpha_{T_0}, \alpha_{T_1}, \alpha_{T_2})$, so that for the context entry with index $j$, both CDF and PARA parameters are initialized as the following array of parameters:
\begin{equation}
    \begin{bmatrix}
c_0^{(0)},\dots,c_{M-1}^{(0)}, \left(\alpha_{T_0}^{(0)}, \alpha_{T_1}^{(0)}, \alpha_{T_2}^{(0)} \right) \\
\vdots \\
c_0^{(j)},\dots,c_{M-1}^{(j)}, \left(\alpha_{T_0}^{(j)}, \alpha_{T_1}^{(j)}, \alpha_{T_2}^{(j)} \right)   \\
\vdots \\
c_0^{(J-1)},\dots,c_{M-1}^{(J-1)}, \left(\alpha_{T_0}^{(J-1)}, \alpha_{T_1}^{(J-1)}, \alpha_{T_2}^{(J-1)} \right)   \\
\end{bmatrix},
\end{equation}
which can be viewed as an context initialization table for $J$ context entries (symbol groups) with symbol length $M$, where the initial CDF entries and the PARA parameters are initialized.

For simplicity, four possible integer values are allowed in AV2 as the PARA parameters which are $\alpha = \{0,1,-1,-2\}$. Specifically, in Equation \ref{eqn:mu_term_with_alpha} setting $\alpha=0$ does not change the speed of adaptation and setting $\alpha$ to $-1$ and $-2$ result in a faster adaptation.

\subsection{Adaptive Transform Coding (ATC)}\label{sec:atc}
ATC is a new coefficient coding scheme in AV2 that introduces a set of coefficient coding rules.
ATC aims to guide the coefficient coding process to more accurately model the statistics of transform coefficients, particularly for low-frequency terms while also reducing the memory footprint of context models.
The ATC design consists of a frequency-aware adaptive region determination with unified scanning and improved context derivation rules for each region, and a simplification of the chroma coefficient coding logic.

\begin{figure}[!b]
\centering
\includegraphics[width=0.5\textwidth]{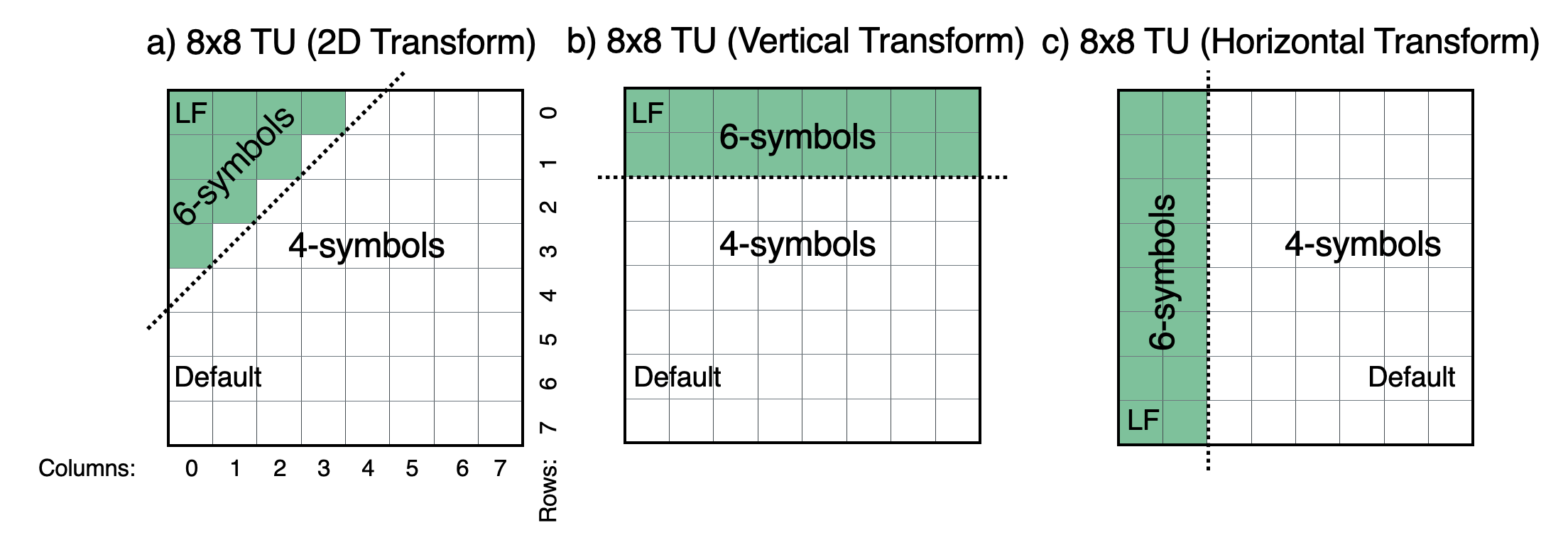}
\caption{Low-frequency and Default Coefficient Regions defined in AV2 for the Luma channel. For chroma, only the DC term (row=0, col=0) is included in the LF region for 2D transforms. For chroma 1D transforms only the first row and column are included.}
\label{fig:atc1}
\end{figure}

\subsubsection{Adaptive Coefficient Coding}\label{sec:atc}
As described in Section~\ref{av1:level_map}, AV1 codes transform coefficient magnitudes using multiple passes: a Base Range (BR), Low Range (LR), and a High Range (HR) pass.

AV2 refines this process by defining two spatial regions within each transform block (TB): a Low-Frequency (LF) region and a Default region which are illustrated in Figure \ref{fig:atc1}.
For 2D transforms, a unified up-right diagonal coefficient scan is applied, and the LF region is defined by the condition $r + c < 4$, where $r$ and $c$ denote the row and column indices within the TB.
Coefficients which do not satisfy the condition ($r + c \geq 4$) fall into the Default region.
The unified scan removes the coefficient scan switching logic from AV1 that depended on the TB dimensions.

For 1D transforms, which use row-wise or column-wise scanning as in AV1, the LF region includes the first two columns for horizontal 1D transforms or first two rows for vertical 1D transforms as shown in Figure~\ref{fig:atc1}b and Figure~\ref{fig:atc1}c.

The LF region allows for more precise context modeling leading to improved arithmetic coding efficiency.
In contrast, the Default region benefits from simplified context derivation rules enhancing entropy coding throughput and faster adaptation due to lesser switching of entropy models.

ATC modifies the coefficient coding syntax in each region for AV2.
In the LF region, the BR level range is extended to cover the level values $\text{BR}_{\text{LF}} = [0, 1, 2, 3, 4, \geq 5)$ using a 6-symbol alphabet.
The final symbol $\geq 5$ serves as an escape character indicating that the actual value may exceed 4 which may trigger another coding pass.
In the Default region, the BR range remains identical to AV1: $\text{BR}_{\text{Default}} = [0, 1, 2, \geq 2)$ including level values $0, 1, 2$ and the escape character for greater $\geq 2$.

\begin{figure}
\centering
\includegraphics[width=0.48\textwidth]{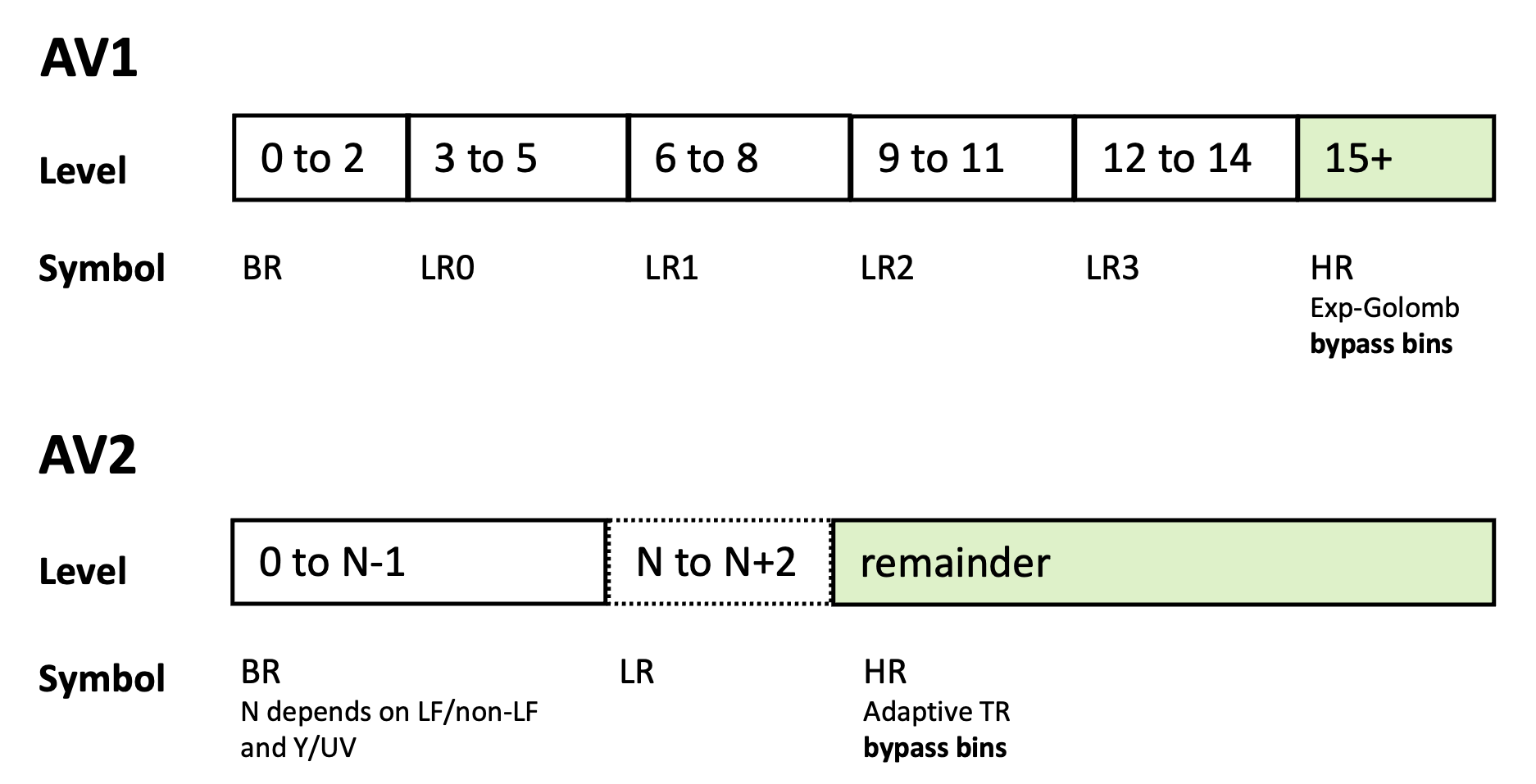}
\caption{AV1 and AV2 coefficient level coding.}
\label{fig:coeff_levels_av1_av2}
\end{figure}

If a coefficient's magnitude equals the escape symbol of the BR range, a subsequent low-range (LR) coding pass is initiated.
In the LF region, the LR pass extends level coding support up to $\geq 8$, while in the Default region it extends to $\geq 6$. Unlike AV1 the LR coding pass in AV2 is performed only once with the levels exceeding these values are encoded in the HR pass using the new adaptive bypass coding method described in Section \ref{sec:adaptive_hr}.
Figure \ref{fig:coeff_levels_av1_av2} summarizes the differences in coding passes between AV1 and AV2 explicitly.

\paragraph{Context Derivation Rules}
Arithmetic coding efficiency is closely tied to accurate context modeling.
AV2 leverages the values of previously decoded coefficients in a local neighborhood to derive a predictive context for each coefficient.
This neighborhood facilitates estimation of the magnitude of the current coefficient allowing selection of the most appropriate probability model.

\begin{figure}[!t]
\centering
\includegraphics[width=0.5\textwidth]{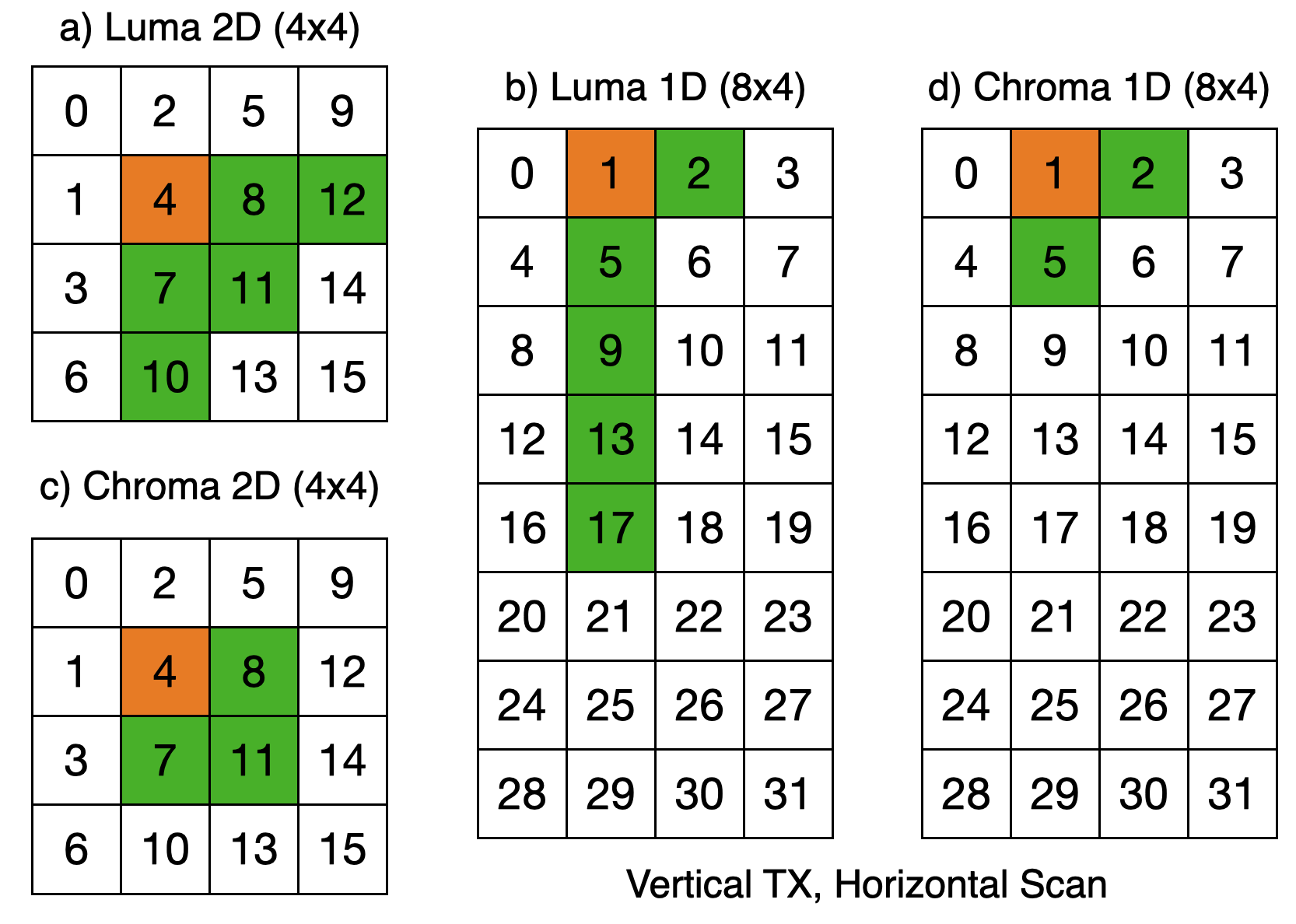}
\caption{Context derivation neighborhoods for luma and chroma channels. (a) 2D luma uses 5 neighbors (green) to derive context for the orange coefficient at position 4. (b) 1D luma uses an inverse-L neighborhood. Chroma uses 3 neighbors for 2D and 2 neighbors for 1D transforms.}
\label{fig:atc2}
\end{figure}

In Figure \ref{fig:atc2}, the neighborhoods used for context derivation are illustrated for luma and chroma channels as well as for 2D and 1D transform types.
For 2D luma transforms, a 5-sample neighborhood (highlighted in green) around the current coefficient (position 4 in orange) is used to compute a local statistic:
\begin{equation}
nstats = \left( \sum_{k} |c_k| + 1 \right) \gg 1
\end{equation}
where $\gg$ denotes the bitwise right-shift operator. This statistic sums the level values of the previously decoded coefficients.
The statistic is then clipped and offset according to the coefficient's position to obtain the context index or a relevant CDF table entry:
\begin{equation}
ctxId = \min(nstat, ~\alpha) + \text{offset}, \quad \alpha \in \{3, 4, 6, 8\}
\end{equation}
Table \ref{tab:atc_br_luma} summarizes the clip value $\alpha$ and offset values for the luma BR coefficients.

\input{tables/atc_table1_br_ctx_luma.tex}
The design allocates more contexts for LF coefficients particularly for the DC and lower-order AC terms, and fewer contexts for default and 1D coefficients.
AV2 also departs from AV1's design by employing distinct and simplified context derivation rules for chroma.
As shown in Table \ref{tab:atc_br_uv} chroma uses a fixed clip of 3 and requires fewer contexts reducing complexity and memory requirements.

\input{tables/atc_table2_br_ctx_chroma.tex}

For the LR coding pass the context index is derived similarly from \texttt{nstats} with rules varying by channel (luma or chroma) and region (LF or Default) as detailed in Table \ref{tab:low_range_context_av2}. Luma uses a clip of 6, while chroma uses a clip of 3. Additional offsets are applied based on coefficient position (e.g., DC vs. non-DC) in the LF region.
Chroma does not have a LR pass in the LF region since BR has more symbol counts.

\input{tables/atc_table3_lr_ctx.tex}

\paragraph{Unified Scanning for 2D Transforms}
AV1 employs different scan orders based on transform block dimensions: up-right diagonal for wide blocks, bottom-left diagonal for tall blocks, and zig-zag for square blocks.
AV2 simplifies this by using a unified up-right diagonal scan for all 2D TBs regardless of TB dimensions.
This reduces complexity by eliminating conditional switching of scan orders based on TB size and coding pass.

\subsection{Adaptive High Range Coding via Truncated Rice}\label{sec:adaptive_hr}
Coefficients whose magnitudes exceed the BR and LR ranges are coded in the high-range (HR) stage using bypass bins.
AV2 replaces AV1 exponential-Golomb HR coding with an \emph{adaptive truncated Rice} (TR) code to better match large-magnitude distributions while bounding the worst-case codeword length.

TR follows Golomb-Rice coding when the unary prefix is short and switches to Exp-Golomb when the prefix becomes long.
This preserves adaptive Rice-parameter selection while keeping the maximum codeword length within 32 bits.
The Rice parameter $m$ is selected from a local context value $ctx$ that tracks recently decoded HR increments:
\begin{equation}
ctx_{t+1} = (ctx_t + \ell_{\text{HR}}) \gg 1,
\end{equation}
initialized to zero per TB. The mapping from $ctx$ to $m$ can be written compactly as
\begin{equation}
m = \min\!\left(6,\, \max\!\left(1,\, \left\lfloor \log_2\!\big(\max(ctx,1)\big) \right\rfloor \right)\right),
\end{equation}
where $\lfloor\cdot\rfloor$ denotes the floor operator.
Given $m$, TR codes the quotient using a unary prefix capped by
\begin{equation}
c_{\max} = \min(m+4, 6).
\end{equation}
Let $q$ denote the decoded unary prefix length. If $q < c_{\max}$, TR reads an $m$-bit remainder (Golomb--Rice).
If $q = c_{\max}$, TR switches to Exp-Golomb coding of order $k=m+1$ for the remaining value.

The decoded HR increment reconstructs the magnitude as
\begin{equation}
\ell = \ell_{\text{BR/LR}} + (\ell_{\text{HR}} \ll s), \quad s =
\begin{cases}
1, & \text{if TCQ is enabled}, \\
0, & \text{otherwise}.
\end{cases}
\end{equation}
In Figure \ref{fig:coeff_levels_av1_av2} the TR logic handles the HR coding-pass in AV2 which replaces the Exp-Golomb coding of AV1.
Truncated Rice coding enables better compression efficiency, particularly at lower QPs and in lossless coding scenarios where precise modeling of large coefficient values significantly improves rate-distortion performance.

\subsection{Forward Skip Coding (FSC)}\label{sec:fsc}
Forward skip coding (FSC) is an AV2 residual-coding mode that improves coefficient-coding efficiency for transform blocks (TBs) coded with the identity transform (IDTX) applied in 2D.
FSC is signaled at the coding-block (CB) level, and all TBs within the CB follow FSC coefficient-coding rules:

\begin{figure}[!t]
\centering
\includegraphics[width=0.85\linewidth]{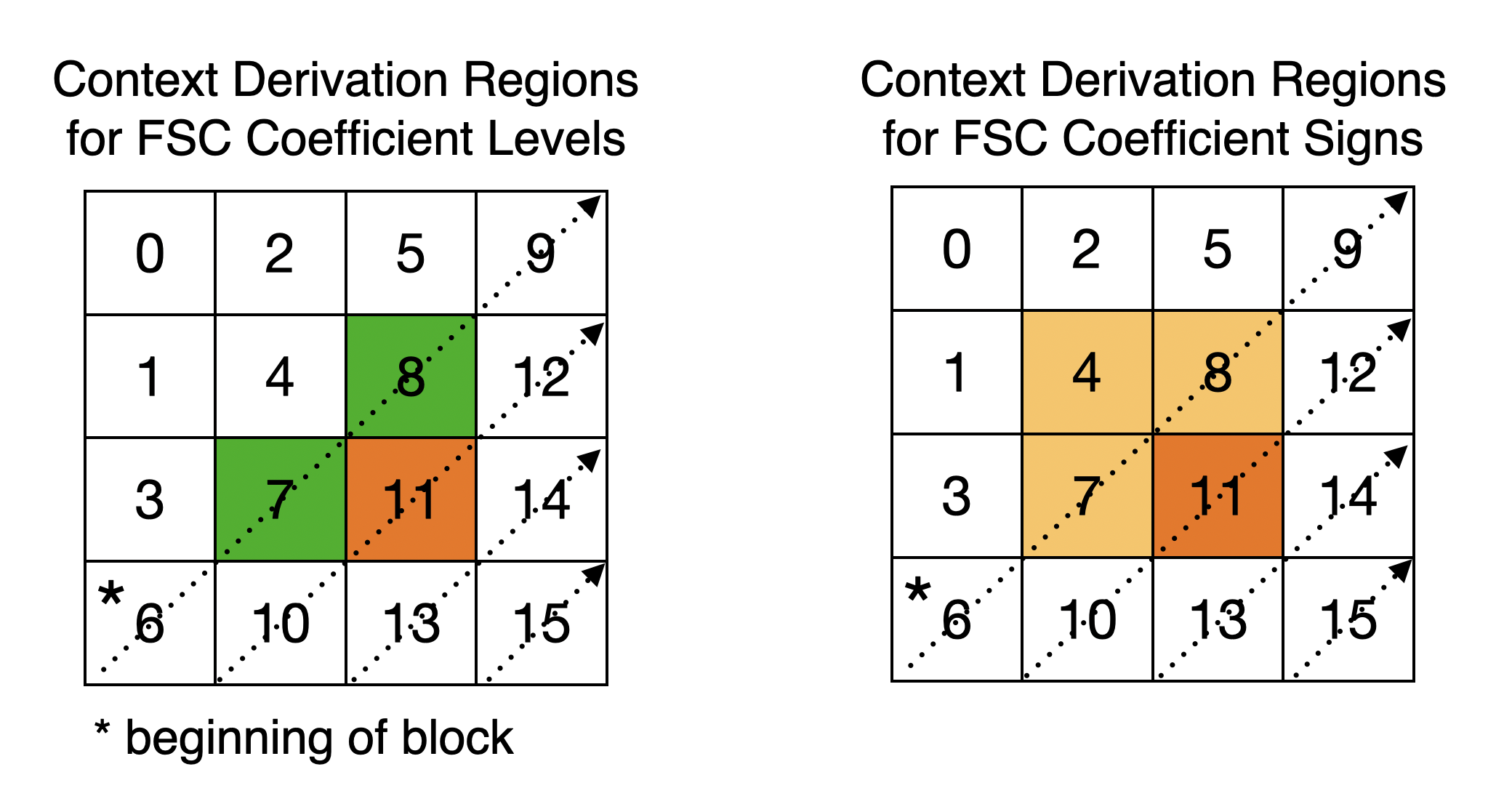}
\caption{Context derivation neighborhoods for FSC: two neighbors (left, above) for level contexts and three neighbors (left, above, above-left) for sign contexts.}
\label{fig:fsc1}
\end{figure}

\begin{itemize}
\item \textit{Beginning-of-block (BOB) signaling:} A first position index referred to as the beginning of block (BOB) is signaled for FSC-coded TBs to indicate the position of the first coded coefficient.
No new syntax is introduced for coding BOB; instead, BOB syntax follows the same logic as EOB (end-of-block) syntax signaling rules in AV2 but uses FSC-specific entropy contexts.
Coefficient coding begins from the BOB location and proceeds in forward scan order until the max coefficient index in a given TB, enabling early skipping of leading zeros and improving entropy throughput.

\item \textit{Coding passes:} Coefficients are parsed in \emph{forward} scan order starting from the BOB position, and magnitudes are coded in three stages aligned with the Default-region syntax in Section~\ref{sec:atc}. BR uses a 4-symbol alphabet $[0,1,2,\ge 2]$; if BR escapes, LR codes $[3,4,5,\ge 5]$; if LR escapes at $\ge 5$, the remaining magnitude is coded in HR using the adaptive bypass method in Section~\ref{sec:adaptive_hr}. Sign information is coded in the final stage together with HR.

\item \textit{Simplified context modeling:} Coefficient context derivation is reduced from 5 neighboring coefficients to just 2 (left and above) for coefficient levels, and 3 neighboring samples (left, above, above-left) for coding the sign information as shown in Fig.~\ref{fig:fsc1}.
The sign and level information are coded using an up-right diagonal scan.
Table~\ref{tab:fsc_1} provides the context numbers and derivation rules for each coding pass: the context derivation for level values (BR and LR passes) follows the previously decoded coefficient samples in the neighborhood with a clip, and a particular offset given the TB size.

\item \textit{TB-level syntax skipping:} For intra blocks coded with FSC, the transform type of all sub-TBs is inferred as IDTX, avoiding TB-level transform-type signaling. For inter-coded blocks, IDTX is explicitly signaled at the TB level, and the signaled IDTX implies FSC.
\end{itemize}

\begin{table}[!t]
\caption{FSC coding passes and context usage.}
\label{tab:fsc_1}
\centering
\resizebox{1\linewidth}{!}{%
\begin{tabular}{|c|>{\raggedright\arraybackslash}p{0.80\linewidth}|c|}
\hline
\textbf{Pass} & \textbf{Level range / context derivation} & \textbf{Contexts} \\
\hline
BR \& LR &
$\mathrm{ctxId}=\mathrm{CLIP}\!\bigl(\mathrm{level}_{\mathrm{left}}+\mathrm{level}_{\mathrm{top}},\,6\bigr)+\mathrm{TX\_SIZE}_{\mathrm{offset}}$
& 21 \\
\hline
HR & adaptive bypass coding (Section~\ref{sec:adaptive_hr}) & 0 \\
\hline
Sign (with HR) &
{\linespread{0.95}\selectfont
$\begin{aligned}
s &= \mathrm{sign}_{\mathrm{left}}+\mathrm{sign}_{\mathrm{top}}+\mathrm{sign}_{\mathrm{top\text{-}left}} \\
\mathrm{ctxId}(s) &= \textstyle
\begin{cases}
5+\delta \protect\footnotemark, & s>2 \\
6+\delta, & s<-2 \\
1+\delta, & s>0 \\
2+\delta, & s<0 \\
\mathrm{TX\_SIZE}_{\mathrm{offset}}, & s=0
\end{cases}
\end{aligned}$}
& 27 \\
\hline
\end{tabular}%
}
\end{table}
\footnotetext{$\delta = 2+\mathrm{TX\_SIZE}_{\mathrm{offset}}$ if $\mathrm{level}_{\mathrm{curr}}>3$; else $\delta=\mathrm{TX\_SIZE}_{\mathrm{offset}}$.}

Overall, FSC provides a low-complexity enhancement for IDTX-coded TBs, with particular benefit for screen and natural content and for near-lossless/lossless operation.
It achieves screen-content gains comparable to the Palette tool in AV2, provides coding gains for natural content, and improves performance in near-lossless and lossless coding cases.

\subsection{Parity Hiding (PH)}\label{sec:parhide}
Parity hiding is a novel entropy coding tool in AV2 that infers the parity of the DC coefficient level in a transform block based on the magnitudes of the coefficients preceding the AC terms in reverse scan order. This enables implicit signaling of the DC parity, thereby reducing the coefficient level value range by half and lowering the number of explicitly coded bits.

PH is applied to transform blocks that use a trigonometric transform in at least one direction (i.e., not used for 2D IDTX). PH is disabled for lossless coding and for chroma blocks. When a transform block has at least four non-zero AC coefficients, the encoder computes the parity of their sum and uses it to implicitly determine the parity of the DC coefficient.

If the preceding-AC coefficients indicate that the DC term should be parity-hidden, the decoder reconstructs the DC coefficient level value as:
\begin{equation}
qcoeff_{DC} = \left( [\text{BR}_{DC} + \text{LR}_{DC} + \text{HR}_{DC}] \times 2 + \text{parity} \right) \times \text{sign}
\end{equation}
where BR, LR, HR indicate the total reconstructed level sum based on Base, Low, and High range coding passes. Otherwise, if the number of non-zero AC coefficients is fewer than four, the DC coefficient is reconstructed conventionally:
\begin{equation}
qcoeff_{DC} = [\text{BR}_{DC} + \text{LR}_{DC} + \text{HR}_{DC}] \times \text{sign}
\end{equation}
The parity term is computed over the non-zero AC terms as:
\begin{equation}
\text{parity} = \left( \sum_i [\text{BR}_{AC_i} + \text{LR}_{AC_i}] \right) \bmod 2
\end{equation}

\subsubsection{Encoder Behavior and Trellis Tuning}
After trellis quantization, the encoder evaluates the cost of possible parity adjustments to match the desired parity without significantly increasing rate. When needed, it selectively perturbs one or more coefficients as an encoder optimization (e.g., adding 1 to a zero or tweaking an existing value by $\pm1$) to align the summed AC parity with the desired hidden parity of the DC coefficient.

This adjustment is guided by a decision table depending on the number of non-zero AC coefficients, as shown in Table~\ref{tab:ph_encoder_detailed}.

\begin{table}[!h]
\caption{Encoder Strategy for Parity Hiding Based on Non-Zero AC Coefficients}
\label{tab:ph_encoder_detailed}
\centering
\setlength{\tabcolsep}{3pt}
\renewcommand{\arraystretch}{1.2}
\begin{tabular}{|c|p{4.3cm}|c|}
\hline
\textbf{\begin{tabular}[c]{@{}c@{}}Non-Zero \\AC Coefficients\end{tabular}} & \textbf{Search Minimal Cost Among} & \textbf{PH State} \\
\hline
3 & Adjust one zero to non-zero & Enabled \\
  & Disable PH for the block     & Disabled \\
\hline
4 & Adjust one zero to non-zero             & Enabled \\
  & Tweak non-zero by $\pm1$               & Enabled \\
  & Adjust parity to be hidden             & Enabled \\
  & Adjust non-zero to zero                & Disabled \\
\hline
$>4$ & Adjust parity of one coefficient       & Enabled \\
     & Adjust parity to be hidden            & Enabled \\
\hline
\end{tabular}
\end{table}

\subsubsection{Context Models}
To preserve coding efficiency, a minimal set of additional context models is introduced for the DC coefficient in the presence of parity hiding. If the parity is hidden, 5 context models are used to code the Base Range (BR) level values, and 7 context models are used to code the Low Range (LR) level values. The context derivation neighborhood is aligned with non-PH regions. Sequence- and frame-level flags control PH enablement.

\begin{table}[!ht]
\caption{Parity Hiding Context Derivation}
\label{tab:ph_context}
\centering
\setlength{\tabcolsep}{3pt}
\renewcommand{\arraystretch}{1.3}
\begin{tabular}{|c|>{\raggedright\arraybackslash}m{5.2cm}|c|}
\hline
\textbf{Pass} & \textbf{Level Range / CTX Derivation} & \textbf{Contexts} \\
\hline
Base Range (BR) &
\makecell[l]{%
$\text{ctxId} = \mathrm{CLIP}\Big((\text{level}_\text{R} + \text{level}_\text{RR} + \text{level}_\text{B} $ \\
$\quad + \text{level}_\text{BB} + \text{level}_\text{RB} + 1 ) \gg 1,\ 4 \Big) $
}
& 5 \\
\hline
Low Range (LR) &
\makecell[l]{%
$\text{ctxId} = \mathrm{CLIP}\Big((\text{level}_\text{R} + \text{level}_\text{B} $ \\
$\quad + \text{level}_\text{RB} + 1) \gg 1,\ 6 \Big)$
}
& 7 \\
\hline
\end{tabular}
\end{table}

\subsection{Quantization Matrices}
Quantization matrices (QMs) are used to weight transform coefficients prior to quantization, reducing visible artifacts by preserving low-frequency energy while allowing higher frequencies to be quantized more aggressively.

In AV1, QMs were largely limited: about 100 kB of tables were defined in the specification (15 levels × 2 planes × all transform sizes). User-defined QMs were not allowed.

AV2 introduces a more flexible framework. AV2 supports user-defined matrices for transform sizes smaller than 8x8, 8x4, and 4x8, and segment-adaptive QMs, allowing application-specific or aggressive scaling at low bitrates. QM parameter signaling for user-defined matrices is further optimized by transmitting only half of symmetric 8×8 matrices and collapsing long sequences of repeated values with a stop symbol.
Additionally, a new open bitstream unit (OBU) has been introduced, which allows per frame updating of the QMs used and user-defined QMs.
Collectively, these improvements make QMs in AV2 more flexible and perceptually effective, improving quality at low bitrates with negligible coding loss.

\subsection{Lossless Coding Improvements}\label{sec:lossless}
To reduce statistical redundancy in the residual domain under a lossless configuration, a residual-block refinement (RBR) method is applied to produce a sparser residual after prediction.
The RBR process is lossless and invertible and the resulting residual serves as the input to the transform and entropy-coding modules.

The refinement has two independent modes: vertical and horizontal refinement modes.
The vertical mode can be expressed in matrix form as:
$
\mathbf{R}' = \mathbf{W}_{v} \times \mathbf{R}
$,
and for horizontal block refinement, the matrix calculation can be formulated as:
$
\mathbf{R}' = \mathbf{R} \times \mathbf{W}_{h}
$.
The original residual block is denoted by $\mathbf{R}$, and its refined counterpart by $\mathbf{R}'$. Vertical refinement is realized by left-multiplying $\mathbf{R}$ with a refinement matrix, which predicts each row from the immediately preceding (upper) row.
Horizontal refinement is realized by right-multiplying $\mathbf{R}$ with a refinement matrix, which predicts each column from the immediately preceding (left) column.
For illustration, consider a $4 \times 4$ block; the corresponding weight matrix is defined as:
\begin{equation*}
\begin{aligned}
W_{v} &=
\small\begin{bmatrix}
1 & 0 & 0 & 0\\
-1 & 1 & 0 & 0 \\
0 & -1 & 1 & 0 \\
0 & 0 & -1 & 1
\end{bmatrix},
\hspace{1mm} %
W_{h} &=
\small\begin{bmatrix}
1 & -1 & 0 & 0\\
0 & 1 & -1 & 0 \\
0 & 0 & 1 & -1 \\
0 & 0 & 0 & 1
\end{bmatrix}.
\end{aligned}
\end{equation*}

For the inverse vertical refinement, the matrix calculation is formulated as:
$
\mathbf{R} = [\mathbf{W}_{v}]^{-1} \times \mathbf{R}',
$
and for the inverse horizontal refinement, the matrix calculation is formulated as:
$
    \mathbf{R} = \mathbf{R}' \times [\mathbf{W}_{h}]^{-1}.
$
The weight matrix are defined as:
\begin{equation*}
\begin{aligned}
[W_{v}]^{-1} &=
\small\begin{bmatrix}
1 & 0 & 0 & 0\\
1 & 1 & 0 & 0 \\
1 & 1 & 1 & 0 \\
1 & 1 & 1 & 1
\end{bmatrix},
\hspace{1mm} %
[W_{h}]^{-1} &=
\small\begin{bmatrix}
1 & 1 & 1 & 1\\
0 & 1 & 1 & 1 \\
0 & 0 & 1 & 1 \\
0 & 0 & 0 & 1
\end{bmatrix}.
\end{aligned}
\end{equation*}

RBR is only allowed for intra blocks and disabled for inter blocks.
Horizontal refinement is applied only when the intra-prediction mode is horizontal, and vertical refinement only when the mode is vertical; no refinement is used for other intra-prediction modes.

RBR results in sparser residuals with smaller magnitude and demonstrates notable improvements with the identity transform and with larger block sizes, whereas the Walsh–Hadamard Transform (WHT) remains advantageous for smooth signal patterns.

\begin{table*}[!t]
\centering
\caption{AOM Common Test Condition (CTC) configurations and corresponding encoder command-line parameters.}
\label{tab:ctc_configs_and_cmds}
\footnotesize
\setlength{\tabcolsep}{4pt}
\renewcommand{\arraystretch}{1.15}
\begin{tabular}{|l|c|c|c|}
\hline
\textbf{Configuration} &
\textbf{Prediction Structure} &
\textbf{Structural Delay} &
\textbf{QP (qindex) Values} \\
\hline\hline

\multirow{2}{*}{\parbox[c]{2.8cm}{\centering\textbf{All Intra (AI)}}} &
Intra-only &
None &
85, 110, 135, 160, 185, 210 \\
\cline{2-4}
&
\multicolumn{3}{p{13.6cm}|}{%
\scriptsize
\texttt{--cpu-used=0 --passes=1 --end-usage=q --qp=\{x\}
--kf-min-dist=0 --kf-max-dist=0
--use-fixed-qp-offsets=1 --deltaq-mode=0
--enable-tpl-model=0 --enable-keyframe-filtering=0
--obu --limit=\{x\}}} \\
\hline\hline

\multirow{2}{*}{\parbox[c]{2.8cm}{\centering\textbf{Random Access (RA)}}} &
Bi-directional inter prediction &
Non-zero &
110, 135, 160, 185, 210, 235 \\
\cline{2-4}
&
\multicolumn{3}{p{13.6cm}|}{%
\scriptsize
\texttt{--cpu-used=0 --passes=1 --lag-in-frames=19
--auto-alt-ref=1 --min-gf-interval=16
--gf-min-pyr-height=4 --gf-max-pyr-height=4
--kf-min-dist=65 --kf-max-dist=65
--use-fixed-qp-offsets=1 --deltaq-mode=0
--enable-keyframe-filtering=0 --enable-tpl-model=0
--end-usage=q --obu --limit=130}} \\
\hline\hline

\multirow{2}{*}{\parbox[c]{2.8cm}{\centering\textbf{Low Delay (LD)}}} &
Single-direction inter prediction &
Zero &
110, 135, 160, 185, 210, 235 \\
\cline{2-4}
&
\multicolumn{3}{p{13.6cm}|}{%
\scriptsize
\texttt{--cpu-used=0 --passes=1 --lag-in-frames=0
--min-gf-interval=16 --max-gf-interval=16
--gf-min-pyr-height=4 --gf-max-pyr-height=4
--kf-min-dist=9999 --kf-max-dist=9999
--use-fixed-qp-offsets=1 --deltaq-mode=0
--enable-tpl-model=0 --end-usage=q --qp=\{x\}
--subgop-config-str=ld
--enable-keyframe-filtering=0 --obu --limit=130}} \\
\hline

\end{tabular}
\end{table*}

\subsubsection{Lossless Transforms}
The lossless mode of AV2 provides significant transform coding improvements over AV1. Specifically, AV2 uses the identity transform with adaptable transform sizes in addition to the $4\times4$ Walsh-Hadamard Transform (WHT).

\paragraph{Transform Scheme for Lossless Coding of Luma}
Lossless coding uses the FSC mode to convey the transform type for luma intra blocks.
The transform type is determined as IDTX when FSC is enabled at the CB level, and as WHT when FSC is disabled.
AV2 also supports the use of larger identity transforms of sizes up to $32 \times 32$.
If a lossless intra block uses FSC, an additional symbol is transmitted to indicate the transform size.
The two permissible transform sizes are $4 \times 4$ and $\min(M,32) \times \min(N,32)$, where $M \times N$ denotes the coding block size.
The combination of the signaled transform size and type uniquely specifies the transform applied to the block.

Luma inter blocks utilize the same set of transforms as luma intra blocks but a different signaling scheme is employed.
Inter blocks do not explicitly signal FSC mode and as in AV1 IDTX is signaled at TB level.
FSC is still used the coefficient coding mode yielding additional coding gains for lossless blocks when IDTX is used.
The transform size ($4 \times 4$ or $\min(M,32) \times \min(N,32)$) is transmitted prior to the transform type (WHT/IDTX), consistent with regular residual coding modes.
The transform type is explicitly signaled only when the transform size is $4 \times 4$, as larger sizes inherently imply the use of IDTX.

\paragraph{Transform Scheme for Lossless Coding of Chroma}
In AV2 the identity transform is also supported for lossless chroma blocks.
The transform type of a chroma block is derived directly from the colocated luma block, while its transform size is restricted to $4 \times 4$.
This design avoids any additional signaling overhead.

\section{Coding Gain Estimates}\label{sec:coding-gains}
The advances in transform and entropy coding design in AV2 provide significant subjective and objective quality improvements over AV1.
Each major coding tool described in this paper underwent multiple revisions during the development cycle and contributes to the overall coding gain targets established by the AOM.
Tool adoption in AV2 is guided by a careful trade-off between compression efficiency and computational complexity.
In this section, we provide estimates of the coding gains attributable to each major tool, as well as cumulative gains, evaluated separately for natural content and screen content.

Compression efficiency is measured using the Bjøntegaard Delta (BD) rate for PSNR and VMAF ~\cite{bjontegaard2001calculation} under the AOM Common Test Conditions (CTC).

\subsection{Common Test Conditions}
The All Intra (AI) configuration evaluates intra coding performance by encoding each frame independently without inter prediction. The Random Access (RA) configuration targets streaming and video-on-demand use cases and employs a hierarchical inter prediction structure with periodic intra frames.Bidirectional prediction is used to improve compression efficiency while maintaining reasonable random access capability. The Low Delay (LD) configuration targets real-time and interactive applications and enforces zero structural frame delay.
Single-direction inter prediction is used by disabling lookahead, frame reordering, and backward references.

Table~\ref{tab:ctc_configs_and_cmds} summarizes the key structural properties of the three CTC configurations, including prediction structure, structural delay constraints, and the QP (qindex) operating points defined by the CTC and the command line arguments used in the AVM reference software.
The sequences evaluated under each configuration are detailed in Appendix~\ref{app:test_sequences}.

\subsection{Lossy Coding Gain Estimates}
Table~\ref{tab:bdrate_summary} reports the BD-rate performance of individual AV2 coding tools for PSNR and VMAF under the AI, RA, and LD configurations.
The reported values correspond to estimated tool contributions obtained either (i) from targeted tool-off experiments under the AVM CTC when a dedicated run-time or high-level control flag is available, or (ii) from coding gains reported in normative proposals when the tool behavior is fully integrated into the reference software.
Due to inter-tool dependencies and subsequent encoder tuning, the individual gains should be interpreted as indicative.
The reported gains are measured relative to the immediately preceding AV2 revision at the time of tool adoption, rather than relative to AV1; as such, direct per-tool comparison against AV1 is not meaningful.

Tools that primarily improve transform-domain energy compaction, including CoreTX enhancements, ATC, and IST, exhibit the largest gains in the AI configuration, where residuals are dominated by intra prediction error.
Among these, IST provides the single largest contribution, delivering substantial improvements for both PSNR and VMAF across all configurations.

Several tools provide consistent gains across both intra- and inter-predicted blocks.
In particular, IST, TCQ, and PH demonstrate stable improvements across AI, RA, and LD configurations, indicating that their effectiveness is not limited to intra residual statistics.
TCQ yields especially strong gains for natural content and perceptual metrics such as VMAF, while its impact on screen content is more limited due to the prevalence of sharp edges and sparse residuals.
PH provides smaller but reliable gains by reducing DC coefficient signaling overhead with negligible complexity impact.

FSC delivers modest gains for natural content but substantial improvements for screen content across all configurations, consistent with its design for IDTX-coded residuals.
CCTX provides smaller yet consistent gains by reducing chroma-plane redundancy with minimal additional complexity.

Overall, the cumulative gains summarized in Table~\ref{tab:bdrate_summary} show average BD-rate gains of approximately $-7.1\%$ PSNR and $-9.7\%$ VMAF in AI, $-4.2\%$ PSNR and $-5.2\%$ VMAF in RA, and $-2.7\%$ PSNR and $-3.4\%$ VMAF in LD for natural content.
These results demonstrate that transform and residual coding innovations account for a significant fraction of the total compression gains achieved by AV2 highlighting the central role of improved energy compaction and coefficient coding efficiency.

\begin{table}[!th]
\centering
\caption{BD-rate results (negative is better) for PSNR and VMAF across AI, RA, and LD configurations for each tool at time of adoption into AV2. The values represent estimated individual tool contributions and do not explicitly account for inter-tool dependencies or subsequent tuning. *CoreTX includes gains from core transform partitioning and data-driven transform (DDT) kernel improvements.}
\label{tab:bdrate_summary}
\renewcommand{\arraystretch}{1.1}
\resizebox{\columnwidth}{!}{%
\begin{tabular}{|l|cc|cc|cc|}
\hline
\multirow{2}{*}{\textbf{Tool}} &
\multicolumn{2}{c|}{\textbf{All Intra}} &
\multicolumn{2}{c|}{\textbf{Random Access}} &
\multicolumn{2}{c|}{\textbf{Low Delay}} \\
\cline{2-7}
 & \textbf{PSNR} & \textbf{VMAF}
 & \textbf{PSNR} & \textbf{VMAF}
 & \textbf{PSNR} & \textbf{VMAF} \\
\hline\hline

CoreTX* (Overall) & $-0.45$ & $-0.60$ & $-0.50$ & $-0.92$ & $-0.44$ & $-0.42$ \\
CoreTX* (SCC)     & $-0.20$ & $-0.88$ & $-0.48$ & $-0.52$ & $-0.18$ & $-0.01$ \\
\hline

ATC (Overall)     & $-0.60$ & $-0.78$ & $-0.25$ & $-0.30$ & $-0.14$ & $-0.19$ \\
ATC (SCC)         & $-0.59$ & $-0.79$ & $-0.27$ & $-0.29$ & $-0.15$ & $-0.13$ \\
\hline

DCTX (Overall)    & $-0.26$ & $-0.33$ & $-0.15$ & $-0.17$ & $-0.15$ & $-0.22$ \\
DCTX (SCC)        & $-0.25$ & $-0.17$ & $-0.16$ & $-0.15$ & $-0.17$ & $-0.21$ \\
\hline

TCQ (Overall)     & $-0.81$ & $-1.49$ & $-0.67$ & $-1.30$ & $-0.27$ & $-0.71$ \\
TCQ (SCC)         & $-0.06$ & $-0.10$ & $+0.13$ & $+0.04$ & $+0.12$ & $-0.32$ \\
\hline

IST (Overall)      & $-3.85$ & $-5.13$ & $-1.76$ & $-2.29$ & $-1.09$ & $-1.24$ \\
IST (SCC)          & $-0.67$ & $-0.05$ & $-0.36$ & $-1.12$ & $-0.17$ & $-0.52$ \\
\hline

PH (Overall)      & $-0.34$ & $+0.08$ & $-0.33$ & $-0.05$ & $-0.24$ & $-0.01$ \\
PH (SCC)          & $-0.31$ & $+0.06$ & $-0.29$ & $-0.05$ & $-0.24$ & $-0.11$ \\
\hline

CCTX (Overall)    & $-0.16$ & $+0.14$ & $-0.17$ & $+0.13$ & $-0.16$ & $-0.22$ \\
CCTX (SCC)        & $-0.20$ & $+0.11$ & $-0.21$ & $+0.14$ & $-0.19$ & $-0.21$ \\
\hline

FSC (Overall)     & $-0.36$ & $-0.03$ & $-0.18$ & $-0.08$ & $-0.12$ & $-0.10$ \\
FSC (SCC)         & $-1.69$ & $-0.17$ & $-1.89$ & $-1.05$ & $-2.17$ & $-1.12$ \\
\hline

PARA (Overall)    & $-0.31$ & $-0.51$ & $-0.18$ & $-0.25$ & $-0.11$ & $-0.29$ \\
PARA (SCC)        & $-0.34$ & $-0.31$ & $-0.27$ & $-0.03$ & $-0.45$ & $-0.58$ \\
\hline\hline

\textbf{Summary (Overall)} & $-7.14$ & $-9.65$ & $-4.19$ & $-5.23$ & $-2.72$ & $-3.40$ \\
\textbf{Summary (SCC)}     & $-4.31$ & $-2.30$ & $-3.80$ & $-2.99$ & $-4.60$ & $-3.19$ \\
\hline
\end{tabular}}
\end{table}

\subsection{Lossless Coding Gain Estimates}
Since reconstructed pixels in the lossless configuration are identical to the original pixels, performance evaluation is based solely on bitrate savings.
The coding gain for the lossless configuration is computed as
\begin{equation}
    Coding\ Gain = \frac{Test_{Bitrate} - Anchor_{Bitrate}}{Anchor_{Bitrate}} \times 100\,\%,
\end{equation}
where negative values indicate bitrate reduction relative to the anchor. Table~\ref{tab:lossless_bdrate_summary} summarizes the lossless coding gains achieved by individual tools under AI, RA, and LD configurations.

\begin{table}[!th]
\centering
\caption{Coding gain results (negative is better) across AI, RA, and LD configurations for each lossless tool at time of adoption into AV2.}
\label{tab:lossless_bdrate_summary}
\renewcommand{\arraystretch}{1.1}
\resizebox{\columnwidth}{!}{%
\begin{tabular}{|l|c|c|c|}
\hline
\textbf{Tool} &
\textbf{All Intra (\%)} &
\textbf{Random Access (\%)} &
\textbf{Low Delay (\%)} \\
\hline\hline

RBR (Overall) & $-1.04$ & $-0.39$ & $-0.30$ \\
RBR (SCC)     & $-0.31$ & $-0.26$ & $-0.02$ \\
\hline

FSC 4x4 (Overall) & $-0.64$ & $-0.09$ & $-0.13$ \\
FSC 4x4 (SCC)     & $-1.82$ & $-1.22$ & $-1.20$ \\
\hline

RBR + FSC 4x4 (Overall) & $-1.79$ & $-0.76$ & $-0.75$ \\
RBR + FSC 4x4 (SCC)     & $-2.61$ & $-1.94$ & $-1.72$ \\
\hline

Lossless Transforms (Overall) & $-3.88$ & $-5.93$ & $-7.00$ \\
Lossless Transforms (SCC)     & $-8.20$ & $-9.51$ & $-9.27$ \\
\hline\hline

\textbf{Summary (Overall)} & $-5.67$ & $-6.69$ & $-7.75$ \\
\textbf{Summary (SCC)}     & $-10.81$ & $-11.45$ & $-10.99$ \\
\hline
\end{tabular}}
\end{table}

\section{Conclusion}
This paper presented a detailed description of the transform and entropy coding design of AV2, demonstrating novel improvements over the AV1 framework.
The paper described the normative design of primary and intra/inter secondary transforms, transform partitioning and signaling, quantization, coefficient coding, and entropy modeling.

On the transform side, AV2 introduces redesigned primary transform kernels, expanded transform partitioning, data-driven transforms, and intra/inter secondary transforms, along with mode- and coefficient-dependent signaling restrictions.
These changes improve energy compaction across a wide range of block sizes, prediction modes, and content characteristics, while reducing implementation complexity through matrix-based formulations and reduced-precision arithmetic.

On the entropy coding side, AV2 extends the AV1 multi-symbol arithmetic coding framework with improved coefficient coding and probability adaptation rules.
Adaptive Transform Coding (ATC), Probability Adaptation Rate Adjustment (PARA), refined high-range coding, and tools such as Trellis Coded Quantization (TCQ), Forward Skip Coding (FSC), Cross-Chroma Component Transforms (CCTX), and Parity Hiding (PH) collectively improve coding efficiency for both natural and screen content, as well as for near-lossless and lossless operation.

Evaluation under the AOM Common Test Conditions shows that the transform and residual coding tools described in this paper contribute a substantial portion of the overall compression gains achieved by AV2 across All Intra, Random Access, and Low Delay configurations.

The BD-rate results show that improvements in transform-domain energy compaction and coefficient coding efficiency account for a significant share of AV2’s bitrate reduction relative to AV1.

Overall, the transform and entropy coding architecture of AV2 represents a comprehensive and carefully balanced evolution of the AV1 design, achieving meaningful compression gains while maintaining decoder throughput, controlling entropy-model memory growth, and preserving practical implementation characteristics for a wide range of deployment scenarios.

\bibliographystyle{IEEEtran}
\bibliography{biblio}

\appendices

\section{Throughput and CDF Memory Impact}\label{sec:throughput_cdf}
In addition to compression efficiency, AV2 development placed strong emphasis on controlling decoder throughput and entropy-model memory growth.
Entropy decoding is a known throughput bottleneck in modern video decoders, and unrestricted growth in context modeling can negatively impact both software and hardware implementations.

Early AVM evaluations showed that the cumulative effect of new entropy coding tools could introduce a modest 12\% decoder throughput regression relative to AV1 in the worst-case scenarios.
This motivated a focused optimization in the Entropy Coding Focus Group of AOM to simplify context modeling and selectively apply bypass coding where the compression benefit was marginal. The effective entropy decoding throughput impact was reduced to approximately 5\% relative to AV1 with a minor coding loss ($<$0.1\% BD-rate).

Entropy coding context (CDF) memory usage was analyzed using a standardized entropy coding memory analyzer \cite{AVM-reference} over the course of development.
The initial AV2 design increased active CDF storage by 2~KBs.
Subsequent simplification reduced this overhead, bringing both RAM and ROM requirements approximately similar to AV1 levels within ~0.5 KBs.

\section{Test Sequences}\label{app:test_sequences}
Table~\ref{tab:ctc_master_sequences} lists the test classes and sequences used for evaluation following the AOM Common Test Conditions.

\begin{table*}[!t]
\centering
\caption{AV2 CTC v8.0 list of video test sequences (4:2:0).}
\label{tab:ctc_master_sequences}
\footnotesize
\renewcommand{\arraystretch}{1.05}
\begin{tabular}{c|c|l|c|c|c}
\toprule
\textbf{Class} & \textbf{No.} & \textbf{Sequence} & \textbf{Resolution} & \textbf{Frame Rate} & \textbf{Bit Depth} \\
\midrule
A1 & 1 & BoxingPractice\_3840x2160\_5994fps\_10bit\_420.y4m & 3840$\times$2160 & 59.94 & 10 \\
A1 & 2 & Crosswalk\_3840x2160\_5994fps\_10bit\_420.y4m & 3840$\times$2160 & 59.94 & 10 \\
A1 & 3 & FoodMarket2\_3840x2160\_5994fps\_10bit\_420.y4m & 3840$\times$2160 & 59.94 & 10 \\
A1 & 4 & Neon1224\_3840x2160\_2997fps.y4m & 3840$\times$2160 & 29.97 & 10 \\
A1 & 5 & NocturneDance\_3840x2160p\_10bit\_60fps.y4m & 3840$\times$2160 & 60 & 10 \\
A1 & 6 & PierSeaSide\_3840x2160\_2997fps\_10bit\_420\_v2.y4m & 3840$\times$2160 & 29.97 & 10 \\
A1 & 7 & Tango\_3840x2160\_5994fps\_10bit\_420.y4m & 3840$\times$2160 & 59.94 & 10 \\
A1 & 8 & TimeLapse\_3840x2160\_5994fps\_10bit\_420.y4m & 3840$\times$2160 & 59.94 & 10 \\
\midrule
A2 & 1 & Aerial3200\_1920x1080\_5994\_10bit\_420.y4m & 1920$\times$1080 & 59.94 & 10 \\
A2 & 2 & Boat\_1920x1080\_5994\_10bit\_420.y4m & 1920$\times$1080 & 59.94 & 10 \\
A2 & 3 & CrowdRun\_1920x1080p50.y4m & 1920$\times$1080 & 50 & 8 \\
A2 & 4 & DinnerSceneCropped\_1920x1080\_2997fps\_10bit\_420.y4m & 1920$\times$1080 & 29.97 & 10 \\
A2 & 5 & FoodMarket\_1920x1080\_5994\_10bit\_420.y4m & 1920$\times$1080 & 59.94 & 10 \\
A2 & 6 & GregoryScarf\_1080x1920p30\_yuv420p10le\_130frames.y4m & 1080$\times$1920 & 30 & 10 \\
A2 & 7 & MeridianTalk\_sdr\_1920x1080p\_5994\_10bit.y4m & 1920$\times$1080 & 59.94 & 10 \\
A2 & 8 & Motorcycle\_1920x1080\_30fps\_8bit.y4m & 1920$\times$1080 & 30 & 8 \\
A2 & 9 & OldTownCross\_1920x1080p50.y4m & 1920$\times$1080 & 50 & 8 \\
A2 & 10 & PedestrianArea\_1920x1080p25.y4m & 1920$\times$1080 & 25 & 8 \\
A2 & 11 & RitualDance\_1920x1080\_5994\_10bit\_420.y4m & 1920$\times$1080 & 59.94 & 10 \\
A2 & 12 & Riverbed\_1920x1080p25.y4m & 1920$\times$1080 & 25 & 8 \\
A2 & 13 & RushFieldCuts\_1920x1080\_2997.y4m & 1920$\times$1080 & 29.97 & 8 \\
A2 & 14 & Skater227\_1920x1080\_30fps.y4m & 1920$\times$1080 & 30 & 10 \\
A2 & 15 & ToddlerFountainCropped\_1080x1080p2997\_yuv420p10le\_130frames.y4m & 1080$\times$1080 & 29.97 & 10 \\
A2 & 16 & TreesAndGrass\_1920x1080\_30fps\_8bit.y4m & 1920$\times$1080 & 30 & 8 \\
A2 & 17 & TunnelFlag\_1920x1080\_5994\_10bit\_420.y4m & 1920$\times$1080 & 59.94 & 10 \\
A2 & 18 & Vertical\_bees\_1080x1920\_2997.y4m & 1080$\times$1920 & 29.97 & 8 \\
A2 & 19 & WorldCup\_1920x1080\_30p.y4m & 1920$\times$1080 & 30 & 8 \\
\midrule
A3 & 1 & ControlledBurn\_1280x720p30\_420.y4m & 1280$\times$720 & 30 & 8 \\
A3 & 2 & DrivingPOV\_1280x720p\_5994\_10bit\_420.y4m & 1280$\times$720 & 59.94 & 10 \\
A3 & 3 & Johnny\_1280x720\_60.y4m & 1280$\times$720 & 60 & 8 \\
A3 & 4 & KristenAndSara\_1280x720\_60.y4m & 1280$\times$720 & 60 & 8 \\
A3 & 5 & RollerCoaster\_1280x720p\_5994\_10bit\_420.y4m & 1280$\times$720 & 59.94 & 10 \\
A3 & 6 & Vidyo3\_1280x720p\_60fps.y4m & 1280$\times$720 & 60 & 8 \\
A3 & 7 & Vidyo4\_1280x720p\_60fps.y4m & 1280$\times$720 & 60 & 8 \\
A3 & 8 & WestWindEasy\_1280x720p30\_420.y4m & 1280$\times$720 & 30 & 8 \\
\midrule
A4 & 1 & BlueSky\_360p25\_v2.y4m & 640$\times$360 & 25 & 8 \\
A4 & 2 & RedKayak\_360\_2997.y4m & 640$\times$360 & 29.97 & 8 \\
A4 & 3 & SnowMountain\_640x360\_2997.y4m & 640$\times$360 & 29.97 & 8 \\
A4 & 4 & SpeedBag\_640x360\_2997.y4m & 640$\times$360 & 29.97 & 8 \\
A4 & 5 & Stockholm\_640x360\_5994.y4m & 640$\times$360 & 59.94 & 8 \\
A4 & 6 & TouchdownPass\_640x360\_2997.y4m & 640$\times$360 & 29.97 & 8 \\
\midrule
A5 & 1 & FourPeople\_480x270\_60.y4m & 480$\times$270 & 60 & 8 \\
A5 & 2 & ParkJoy\_480x270\_50.y4m & 480$\times$270 & 50 & 8 \\
A5 & 3 & SparksElevator\_480x270p\_5994\_10bit.y4m & 480$\times$270 & 59.94 & 10 \\
A5 & 4 & Vertical\_Bayshore\_270x480\_2997.y4m & 270$\times$480 & 29.97 & 8 \\
\midrule
B1 & 1 & CosmosTreeTrunk\_2048x858p24.y4m & 2048$\times$858 & 24 & 8 \\
B1 & 2 & EuroTruckSimulator2\_1920x1080p60.y4m & 1920$\times$1080 & 60 & 8 \\
B1 & 3 & GlassHalf\_1920x1080p24.y4m & 1920$\times$1080 & 24 & 8 \\
B1 & 4 & Life\_1920x1080p30.y4m & 1920$\times$1080 & 30 & 8 \\
B1 & 5 & Minecraft\_1920x1080p60.y4m & 1920$\times$1080 & 60 & 8 \\
B1 & 6 & Sniper\_1920x1080p30.y4m & 1920$\times$1080 & 30 & 8 \\
B1 & 7 & SolLevanteDragons\_1920x1080p24\_10bit.y4m & 1920$\times$1080 & 24 & 10 \\
B1 & 8 & SolLevanteFace\_1920x1080p24\_10bit.y4m & 1920$\times$1080 & 24 & 10 \\
B1 & 9 & StarCraft\_1920x1080p60.y4m & 1920$\times$1080 & 60 & 8 \\
B1 & 10 & Witcher3\_1920x1080p60.y4m & 1920$\times$1080 & 60 & 8 \\
\midrule
B2 & 1 & BigBuckBunnyStudio\_1920x1080p60\_10bit.y4m & 1920$\times$1080 & 60 & 10 \\
B2 & 2 & Debugging\_1920x1080p30.y4m & 1920$\times$1080 & 30 & 8 \\
B2 & 3 & MissionControlClip1\_1920x1080p60\_10bit.y4m & 1920$\times$1080 & 60 & 10 \\
B2 & 4 & MobileDeviceScreenSharing\_1078x2220p15.y4m & 1078$\times$2220 & 15 & 8 \\
\bottomrule
\end{tabular}
\end{table*}

\end{document}

%% file: tables/atc_table4_st.tex
\begin{table}[!t]
\renewcommand{\arraystretch}{1.3}
\caption{Memory requirements and worst case computational complexity of ST}
\label{tab:st_memory}
\centering
\resizebox{\linewidth}{!}{%
\begin{tabular}{|c|c|c|c|}
\hline
\textbf{Transform Block Size} & \textbf{Kernel Dimension} & \textbf{Size (KB)} & \textbf{Worst-case Mults/pixel} \\
\hline
$< 8\times8$     & $8\times16$     & 5.25         & 8 \\
\hline
$\geq 8\times8$   & $32\times48$     & 42.75         & 15 \\
\hline
\end{tabular}
}
\end{table}

%% file: tables/atc_table1_br_ctx_luma.tex
\begin{table}[!b]
\renewcommand{\arraystretch}{1.3}
\caption{AV2 Luma Base Range Context Derivation}
\label{tab:atc_br_luma}
\centering
\resizebox{\linewidth}{!}{%
\begin{tabular}{|c|c|c|c|c|}
\hline
\textbf{Region} & \textbf{TX Type} & \textbf{Condition} & \textbf{Context Formula} & \textbf{CTX Range} \\
\hline
\multirow{5}{*}{LF} 
  & 2D & Row + Col == 0 & $ \min(nstats,\ 8) $ & 0--8 \\
  & 2D & Row + Col $<$ 2 & $ \min(nstats,\ 6) + 9 $ & 9--15 \\
  & 2D & Else & $ \min(nstats,\ 4) + 16 $ & 16--20 \\
  & 1D & Row == 0 $\lor$ Col == 0 & $ \min(nstats,\ 6) + 21 $ & 21--27 \\
  & 1D & Row == 1 $\lor$ Col == 1 & $ \min(nstats,\ 4) + 28 $ & 28--32 \\
\hline
\multirow{4}{*}{Default} 
  & 2D & Row + Col $<$ 6 & $ \min(nstats,\ 4) $ & 0--4 \\
  & 2D & Row + Col $<$ 8 & $ \min(nstats,\ 4) + 5 $ & 5--9 \\
  & 2D & Else & $ \min(nstats,\ 4) + 10 $ & 10--14 \\
  & 1D & Any & $ \min(nstats,\ 4) + 15 $ & 15--19 \\
\hline
\end{tabular}
}
\end{table}

%% file: tables/atc_table2_br_ctx_chroma.tex
\begin{table}[!b]
\renewcommand{\arraystretch}{1.3}
\caption{AV2 Chroma Base Range Context Derivation}
\label{tab:atc_br_uv}
\centering
\resizebox{\linewidth}{!}{%
\begin{tabular}{|c|c|c|c|}
\hline
\textbf{Region} & \textbf{TX Type} & \textbf{Context Formula} & \textbf{CTX Range} \\
\hline
\multirow{2}{*}{LF}
  & 2D & $ \min(nstats,\ 3) + \text{V}_\text{offset} $ & 0--7 \\
  & 1D & $ \min(nstats,\ 3) + 8 $ & 8--11 \\
\hline
\multirow{2}{*}{Default}
  & 2D & $ \min(nstats,\ 3) + \text{V}_\text{offset} $ & 0--7 \\
  & 1D & $ \min(nstats,\ 3) + 8 $ & 8--11 \\
\hline
\end{tabular}

}
\vspace{1em}\\
\footnotesize\textbf{Note:} Formulas and index ranges shown for Plane U. For Plane V, an offset of $\text{V}_\text{offset} = +4$ is applied in the Low-Frequency 2D and Default regions.
\end{table}

%% file: tables/atc_table3_lr_ctx.tex
\begin{table}[!b]
\renewcommand{\arraystretch}{1.3}
\caption{Low Range Context Index Derivation Rules for AV2 Coefficient Coding}
\label{tab:low_range_context_av2}
\centering
\resizebox{\linewidth}{!}{%
\begin{tabular}{|c|c|c|c|}
\hline
\textbf{Region (Plane)} & \textbf{Condition} & \textbf{Context Formula} & \textbf{CTX Range} \\
\hline
LF (Luma)     & $c = 0$     & $ctx = \min(nstats,\ 6)$         & 0--6 \\
              & $c > 0$     & $ctx = \min(nstats,\ 6) + 7$     & 7--13 \\
\hline
Default (Luma)& any         & $ctx = \min(nstats,\ 6)$         & 0--6 \\
\hline
Default (Chroma) & any      & $ctx = \min(nstats,\ 3)$         & 0--3 \\
\hline
\end{tabular}
}
\vspace{0.5em}\\
\footnotesize\textbf{Note:} Chroma coefficients in the LF region do not use Low Range (coeff\_br) coding.
\end{table}